\documentclass[acmsmall, nonacm]{acmart}

\usepackage{subcaption}
\usepackage{cleveref}
\usepackage[subtle]{savetrees}

\usepackage[disable,textsize=scriptsize,backgroundcolor=green]{todonotes}
\newcommand\ml[1]{\todo{{\bf ML}: #1}} 
\newcommand\ms[1]{\todo{{\bf MS}: #1}} 
\newcommand\danh[1]{\todo{{\bf DH}: #1}} 
\newcommand\tm[1]{\todo{{\bf TM}: #1}} 

\usepackage{soul} 

\newcommand{\rev}[1]{#1}

\AtBeginDocument{%
  \providecommand\BibTeX{{%
    \normalfont B\kern-0.5em{\scshape i\kern-0.25em b}\kern-0.8em\TeX}}}

\setcopyright{acmlicensed}
\copyrightyear{2021}

\acmJournal{PACMHCI}
\acmYear{2021} 
\acmNumber{CSCW} 
\acmPrice{15.00}

\begin{document}

\title[The Challenge of Variable Effort Crowdsourcing and How Visible Gold Can Help]{The Challenge of Variable Effort Crowdsourcing and How \\Visible Gold Can Help}

\author{Danula Hettiachchi}\authornote{This work was performed while Danula Hettiachchi was at Amazon.}
\email{danula.hettiachchi@rmit.edu.au}
\orcid{1234-5678-9012}
\affiliation{%
  \institution{School of Computing Technologies, RMIT University}
  \city{Melbourne}
  \country{Australia}
}

\author{Mike Schaekermann}
\email{mschaeke@amazon.com}
\affiliation{%
  \institution{Amazon}
  \city{Toronto}
  \state{Ontario}
  \country{Canada}
}

\author{Tristan McKinney}
\email{tristamc@amazon.com}
\affiliation{%
  \institution{Amazon}
  \city{Palo Alto}
  \state{California}
  \country{USA}
}

\author{Matthew Lease}
\email{ml@utexas.edu}
\affiliation{%
  \institution{Amazon}
  \city{Seattle}
  \state{Washington}
  \country{USA}
}
\affiliation{%
  \institution{School of Information, University of Texas at Austin}
  \city{Austin}
  \state{Taxas}
  \country{USA}
}



\newcommand{\acite}[1]{{\protect\NoHyper\citeauthor{#1}}~\cite{#1}\protect\endNoHyper}
\newcommand{\eg}{\textit{e.g.,}}
\newcommand{\ie}{\textit{i.e.,}}


\begin{abstract}
We consider a class of {\em variable effort} human annotation tasks in which the number of labels required per item can greatly vary (e.g., finding all faces in an image, named entities in a text, bird calls in an audio recording, etc.). In such tasks, some items require far more effort than others to annotate. Furthermore, the per-item annotation effort is not known until after each item is annotated since determining the number of labels required is an implicit part of the annotation task itself. On an image bounding-box task with crowdsourced annotators, we show that annotator accuracy and recall consistently drop as effort increases. We hypothesize reasons for this drop and investigate a set of approaches to counteract it. Firstly, we benchmark on this task a set of general {\em best-practice} methods for quality crowdsourcing. Notably, only one of these methods actually improves quality: the use of {\em visible gold} questions that provide periodic feedback to workers on their accuracy as they work. Given these promising results, we then investigate and evaluate variants of the visible gold approach, yielding further improvement. Final results show a \tm{Should we instead phrase this as a 21\% reduction in error?}7\% improvement in bounding-box accuracy over the baseline. We discuss the generality of the visible gold approach and promising directions for future research.

\end{abstract}





\maketitle

\section{Introduction}

Annotations (aka labels) provide the basis for training and testing supervised learning models. Consequently, ensuring the quality of annotations is important, especially in a crowdsourced setting with remote, inexpert annotators. While quality assurance for crowdsourcing is well-studied, relatively little work has studied {\em variable effort} annotation tasks in which the number of labels required per item can greatly vary. Examples might include labeling all faces in an image, named entities in a text, or bird calls in an audio recording. Because the number of instances to label per item can greatly vary, some items require far more effort than others to annotate. Moreover, because there is typically no natural upper-bound on the number of instances present, some individual items may require enormous effort. Finally, the annotation effort required for each item is not known until after it is annotated since determining the number of labels required is an implicit part of the annotation task itself.

In this paper, we first conceptualize the notion of variable effort tasks and how they differ from more typical annotation tasks. For example, such annotation tasks are implicitly two-step: searching the item for all instances matching a target type (e.g., ``face''), then applying a labeling operation (e.g., bounding box) to each matching instance. With labeling effort proportionate to the size of search results, the variable size of search results is the key challenge. This framing also helps us relate variable effort tasks to a wider class of annotation search tasks~\cite{Kutlu20-jair}.

Next, we empirically investigate the specific variable effort labeling task of {\em object detection}: finding and localizing human faces in Open Images~\cite{OpenImages} (via bounding boxes). Whereas many prior studies on object detection report results on simpler datasets having only a few objects per image, our dataset includes as many as 14 faces per image. Our results show that crowdsourced annotator accuracy and recall on Mechanical Turk (MTurk) drops markedly as the number of faces per image increases. We hypothesize a set of key underlying issues leading contributing to this reduced quality: inconsistency of worker experience, the potential for high cognitive load, and ineffective incentive design. 

To address these issues, we adapt and assess a set of general {\em best-practice} methods for quality crowdsourcing: financial incentives, workflow design, and visible gold (i.e., questions that provide periodic feedback to workers on their accuracy as they work). We implement five specific approaches: variable pay per instance, post-task bonuses, task decomposition, iterative improvement, and in-task visible gold with uniform frequency. Notably, only visible gold improves quality.

Motivated by this finding, we further explore the design space for effective use of visible gold questions in variable effort labeling tasks.
While prior work shows that visible gold can improve data quality~\cite{le2010ensuring,gadiraju2015training}, many questions remain. How should we present feedback for variable effort labeling tasks like object detection? What is the optimal strategy to issue visible gold questions? How can the effect of visible gold be strengthened by quality-related consequences (i.e., warnings and bonuses)?
We explore different variants of visible gold task designs. We find that combining both upfront and regular testing sustains data quality significantly better than upfront or regular testing alone. Moreover, imposing quality-related consequences yields further improvement. Our final variant of visible gold integrates dynamic testing with tier-based consequences and significantly outperforms all other approaches.

{\bf Contributions.} We make three primary contributions in this work:
\begin{enumerate}
    \item We conceptualize a class of \textit{variable effort} human annotation tasks. We identify a unique set of data quality challenges they present, along with an empirical analysis of these challenges in the context of object detection.

    
    \item We \textit{systematically evaluate existing methods} to address these challenges and show that providing in-task feedback through \textit{visible gold} significantly outperforms various other baselines, including approaches that adjust pay according to effort or that standardize effort at constant pay.
    
    
    \item We contribute an in-depth analysis of different visible gold variants investigating \textit{issuance patterns} and \textit{consequences} for workers. Based on these investigations, we propose and evaluate an \textit{improved visible gold design} that significantly increases bounding box accuracy by 5.7\% compared to a basic visible gold variant and by 7.5\% compared to a baseline without visible gold.
\end{enumerate}

\section{Related Work}

\subsection{Financial Incentives and Crowd Work}
\label{sec:payment}

Financial incentives can influence work in various ways: who chooses to accept work, how much work they perform, and the quality of work they produce. \citet{vaughan2017making} presents a valuable, succinct review of related work in this area. Early work suggested quality was not impacted by payment \cite{mason2009financial, buhrmester2011amazon,Grady10}. In some cases \cite{buhrmester2011amazon,Grady10}, the difference in payments may have been too low to influence behavior. \citet{mason2009financial} hypothesized an {\em anchoring effect}, with workers' sense of fair payment anchored by whatever was offered. \citet{ipeirotis-2011} reports a similar finding. 

Whereas the early studies used crowdsourcing tasks that were relatively easy to perform, \citet{ho2015incentivizing} and \citet{Ye2017} instead studied ``effort-responsive tasks'' in which workers could improve output via more time or effort, and did see the quality improve with financial incentives. \citet{yin2016predicting} find that while very engaged or un-engaged workers appear insensitive to price, more middling workers improve quality with financial incentives.

\acite{horton2010labor} frame the issue wrt.\ the economics notion of {\em reservation wage}: ``...the minimum wage a worker is willing to accept ...for performing some task; it is the key parameter in models of labor supply.'' Thus as pay decreases, it could fail to match more workers' reservation thresholds and thus potentially bias the sample of workers who choose to perform the task. 
However, Horton and Chilton find mixed evidence for worker behavior conforming to predictions of the rational model: ``workers are clearly sensitive to price but insensitive to variations in the amount of time it takes to complete a task.'' 


While MTurk's pay-per-task pricing model is familiar in crowdsourcing research, this model encourages work efficiency, but risks rushed work since worker earnings can be increased by completing more tasks in less time. An alternative pricing model is hourly pay. Both \citet{Mankar17-hcomp} and \citet{whiting2019fair} proposed technical approaches making it easier for requesters to offer hourly pay jobs on MTurk. Some commercial vendor workforces\footnote{\url{https://aws.amazon.com/sagemaker/groundtruth/pricing/}} also set fixed hourly pay rates. For example, Amazon SageMaker GroundTruth's popular vendor iMerit\footnote{\url{https://aws.amazon.com/marketplace/pp/B07DK37Q32}} charges \$6.12/hour per worker. While hourly pay has the potential to discourage rushed work, since all time worked is compensated, a vendor workforce may still operate internally on a call-center model, where workers may have productivity quotas that similarly encourage them to work efficiently (to enable the vendor to provide competitive pricing and ensure profitability).

\subsection{Workflow Designs}

A crowdsourcing task workflow defines how a single task is organized into a set of HITs that can be completed by one or more workers. Literature shows that we can improve the data quality by adopting a suitable workflow for each task. 
In this work, we are particularly interested in workflow designs that can be used to standardize the task effort in variable effort tasks. 

Prior work highlights two main workflow paradigms, iterative and parallel~\cite{goto2016understanding}. In {\em iterative} workflows, we present the same task to multiple workers in a sequential manner where workers could see previous workers' responses. \acite{little2010exploring} shows that an iterative workflow can improve the average data quality in writing and brainstorming tasks. However, the paper highlights that work produced through parallel workflows could still yield individual responses with higher quality. In a translation task, \acite{Ambati2012Collaborative} shows that a 3-phased iterative workflow can achieve higher quality than a baseline that gathers individual translations from 5 workers for each item. Iterative workflows can also allow us to engage workers with different expertise levels at each iteration~\cite{Ambati2012Collaborative}.

{\em Parallel} workflows aim to get multiple workers to work on parts of the task at the same time~\cite{goto2016understanding}. Parallel work can be on the same task unit (\ie~obtaining multiple answers for the same unit) or smaller sub-tasks obtained through task decomposition.
Find-Fix-Verify~\cite{Bernstein2010Soylent} is a specific workflow pattern that facilitates task decomposition through the initial find step and works well for writing tasks such as proofreading, formatting, and shortening text~\cite{Bernstein2010Soylent}. Prior work by~\acite{kittur2011crowdforge} proposes a framework for decomposing complex crowd tasks. It shows that in a writing task, articles produced through task decomposition received higher ratings and had lower variability than individual-produced articles. Recent work has also investigated how to optimally decompose a task into atomic sub-tasks considering the desired reliability and cost~\cite{Tong2018SLADE:Crowdsourcing}.

Often a combination of iterative and parallel elements can be used to create a more versatile workflow.
Other notable work includes tools that can help visualize and manage complex crowdsourcing workflows~\cite{kittur2012crowdweaver}, allow workers to create workflows~\cite{Kulkarni2012Collaboratively}, and optimize workflows~\cite{dai2013pomdp}.


\subsection{Gold Standard Questions}

The use of gold standard questions (also known as control or gold questions) is a fundamental and widely used quality control mechanism in crowdsourcing~\cite{Daniel2018}.
By injecting gold standard questions and evaluating responses, requesters can accurately measure worker performance~\cite{huang2013enhancing}.

Prior research has investigated how to include gold standard questions within a crowdsourcing job in a systematic way.
\acite{liu2013scoring} predict the optimum number of gold questions to include for estimation tasks such as estimating the price of a product.
The paper concludes that when using a two-stage estimation (\ie~estimate the worker quality only using gold data), the number of control questions should be equal to the square root of the number of labels provided by the worker as a rule of thumb.
Recent work has also explored more dynamic approaches that leverage gold standard questions to select tasks for workers such that overall accuracy is maximized~\cite{Bragg2016OptimalWorkers,Fan2015,Khan2017CrowdDQS}. However, the problem of utilizing and assigning gold standard questions has been mainly investigated in the context of multiple-choice questions, and some solutions are not generalizable across different task types. In particular, much of the previous work has relied on worker accuracy estimation models that only work with binary outcomes or multiple-choice questions.

On the one hand, using a small pool of gold standard questions can lead to problems when gold questions are repeated and flagged by workers~\cite{checco2018all,checco2020adversarial}.
On the other hand, crowdsourcing is typically used for problems for which sourcing ground truth data is not straightforward.
Thus, creating good gold data at scale and at a low cost is essential for implementing gold standards.
\acite{oleson2011programmatic} propose a programmatic approach to generate gold standard data.
This study indicates that programmatic gold can increase the gold per question ratio, allowing for high-quality data without increased costs.


As opposed to creating gold questions prior to the label collection, we can also iteratively validate selected answers using experts.
For example, \acite{hung2015minimizing} investigate classification tasks and proposes a probabilistic model that can find the most beneficial answer to validate in terms of result correctness and detection of faulty workers.
Reliable and high-quality gold data can also be generated by using domain experts~\cite{hara2013combining}.



\subsection{Visible Gold}

Typically, workers cannot distinguish between a regular question and a gold standard question.
Answers received for gold questions are used to estimate the worker quality in the post-processing step or during run-time.
However, gold standard questions can also be used to provide training and feedback to workers~\cite{le2010ensuring,gadiraju2015training,doroudi2016toward}.

Research shows that providing feedback can enhance data quality in crowdsourcing.
\acite{Dow2012} report that both self-assessment and external expert feedback can improve crowd work quality.
The study highlights that workers who receive external assessments tend to revise their work more~\cite{Dow2012}.
Similarly, feedback from peers in organized worker groups can help workers achieve high output quality~\cite{whiting2017crowd}.
In a peer-review setup, the process of reviewing others' work has also been shown to help workers elevate their own data quality~\cite{Zhu2014}.
While peer and expert feedback can improve data quality, it is difficult to achieve the timeliness that is critical for implementing a feedback system at scale\tm{Is there any evidence that immediate feedback is more useful than delayed feedback? If so, that would be a useful point here.}.
\rev{In addition to feedback on work, workers could also benefit from learning opportunities on how to effectively use the tools and their related metrics~\cite{Savage2020BecomingWorkers}.}

From prior research by \acite{Dow2012}, we can identify three key aspects of feedback for crowd work.
`Timeliness' is how quickly the worker receives the feedback in either a synchronous or asynchronous fashion.
`Specificity' is the level of detail in the feedback, ranging from a binary decision (\eg~approve, reject) to template-based structured feedback to detailed task-specific feedback.
Finally, the `Source' of the feedback could be the requester, experts, peer workers, or the worker him- or herself.

A dedicated training phase where workers complete several training tasks and receive feedback until they reach the desired quality level has also been shown to be effective in crowd tasks that involve complex tools and interfaces~\cite{park2014toward}.
Prior work also shows that training or feedback can also introduce a bias due to the specific examples selected for the training/feedback step~\cite{le2010ensuring}.
Other work uses feedback to clarify ambiguous task instructions as opposed to improving the quality of work.
For instance, \acite{manam2018wingit} propose a Q\&A and Edit functionality that can be used by workers to clarify and improve task instructions or questions.

{\em Visible gold} questions allow us to provide feedback while testing for work quality.
\acite{le2010ensuring} show that in a relevance categorization task, a uniform distribution of labels in visible gold standard data produces optimal peaks when considering individual worker precision, as well as majority voting aggregated results.
Their study includes a dedicated pre-task training phase to qualify for the task.
Visible gold questions are inserted based on a simple ratio where workers encounter 1 visible gold question for every 4 questions.
Workers are also blocked from continuing on a task if their accuracy is low.
Before being blocked, workers receive a warning that their accuracy is too low and that they should reread the instructions to correct mistakes.

\acite{gadiraju2015training} test with two training methods with visible golds.
In implicit training, workers are provided training when they provide erroneous responses to gold questions, and in explicit training, workers are required to go through a training phase before they attempt to work on the task itself.
The results indicate that training provides a 5\% performance gain and 40\% time gain across 4 task types (Information Finding, Spam detection, Sentiment Analysis, Image transcription).
However, the experiment setup doesn't define a specific gold injection strategy for implicit training. Instead, it considers all questions as gold.
Using complex web search challenges as the task, \acite{doroudi2016toward} also show that providing expert examples upfront is an effective form of training.
\ms{For related work in general and this section in particular, we should highlight how this paper expands beyond or is different from existing work.}


\subsection{Bounding Box Annotation}

Early improvements to object detection include improvements to the crowdsourcing task workflow.
Object annotation workflow proposed by \acite{su2012crowdsourcing} entails three steps.
First, a worker draws a bounding box around a single object instance.
Second, another worker verifies the drawn box.
Third, a different worker determines if there are additional instances of the object class that need to be annotated. The paper reports that 97.9\% of images are correctly covered with bounding boxes.

Other approaches use computer vision methods to generate bounding boxes during the annotation process~\cite{papadopoulos2016we,adhikari2020iterative,adhikari2018faster,russakovsky2015best}.
Prior work by \acite{papadopoulos2016we} using an accept/reject decision could achieve high-quality results comparable to standard manual annotation.
Similarly, \acite{adhikari2020iterative} propose a semi-automated batch-wise method where a subset of images are annotated and then used to train an object detection model that can generate bounding boxes for the remaining images.
As the last step, generated annotations go through a manual verification where workers add/remove boxes as required.
This method can reduce the manual effort by up to 75\%. 

Literature has also investigated how we could use different annotation strategies instead of the standard way of drawing a bounding box through click and drag interactions.
For instance, bounding boxes could be auto-generated by asking workers to annotate four edge points (points belonging to the top, bottom, left-most, and right-most parts) of the object~\cite{papadopoulos2017extreme}. 
A similar approach uses a single point that corresponds to the center of the target object as opposed to four edge points~\cite{papadopoulos2017training}.

A key challenge in comparing our empirical results vs.\ those reported in prior studies is that they tend to report on datasets having few objects per image on average: 2.5 for PASCAL VOC 2007 (used by~\cite{papadopoulos2016we,papadopoulos2017extreme}), 2.4 for 2012 (used by~\cite{adhikari2020iterative,papadopoulos2017extreme}) and 1.5 for ImageNet (used by~\cite{su2012crowdsourcing}) datasets~\cite{liu2020deep}.
So while \citet{papadopoulos2017extreme} report 88\% mIoU annotation quality on PASCAL VOC 2017, this is a much easier task than ours. In contrast, \acite{russakovsky2015best} report 7 objects per image on average (similar to us), but they do not report annotator mIoU. 

\section{The Challenge of Variable Effort Labeling Tasks}

\subsection{Defining Variable Effort Labeling Tasks}
\label{sec:defining}

While crowdsourced annotation is well-studied, {\em variable effort} tasks present three key challenges vs.\ more typical labeling tasks: inconsistent worker experience, the potential for high cognitive load, and effective incentive design. With regard to inconsistent experience, workers may implicitly expect all task instances to require comparable effort. Highly varying effort requirements across instances would violate such an expectation and could induce surprise or frustration. Secondly, as the cognitive load becomes excessive (e.g., labeling 1000 faces in a single image of a crowd), workers may not only be frustrated but naturally struggle to complete the task accurately. As for incentive design, the typical task-based pricing model ala MTurk assumes that all task instances are compensated at the same fixed rate. Since more effortful instances require more time to complete (accurately), this equates to a lower effective earning rate for workers. These challenges, taken separately and especially together, can have various negative impacts.  Workers may choose not to accept a task or quickly abandon it. They might complete easy instances but skip over more effortful ones. They may fail to deliver quality work due to demanding cognitive load or simple lack of effort.

\urldef\waldourl\url{https://en.wikipedia.org/wiki/Where%27s_Wally%3F}
Such tasks exemplify the applicability of rationales to a broad class of {\em Where's Waldo?}~\cite{waldo} search problems of determining whether or not a given item contains entities of interest (e.g., does Waldo appear in a given image or video clip, do we hear his voice in a given audio recording, is he discussed in a given text, etc.). The larger the item, the greater the problem searching it. For example, imagine annotating all trees in massive satellite or aerial imagery, requiring annotators to zoom and pan around images. The search problem may be explicit -- e.g., {\em does an audio clip contain a bird call?} -- or implicit -- e.g., {\em rate a product from its description}, where the primary task is to rate the item but the annotator must search the item for evidence to support their rating decision. 

Framing this search problem lets us relate variable effort annotation tasks to a large body of related work mobilizing the crowd for distributed search of large search spaces. Classic examples include the search for extraterrestrial intelligence (SETI@Home) \cite{setiathome}, for Jim Gray's sailboat \citep{jimgray} 
or other missing people \citep{wang2009human}, for DARPA's red balloons \citep{tang2011reflecting}, for astronomical events of interest \citep{lintott2008galaxy}, for endangered wildlife \citep{rosser2019crowds} or bird species \citep{kelling2013human}, etc. \citet{attenberg2011beat} asked the crowd to find examples on which classifiers erred. Across such examples, what is being sought must be broadly recognizable so that the crowd can accomplish the search task without the need for subject matter expertise \citep{Kinney:2008:EDE:1458082.1458160}. Whereas the works above involve searching for an entity across domain instances, with variable effort labeling tasks, the challenge is searching within each instance for matching entities.


There is limited prior work that examines how crowd work quality can vary when attempting tasks that involve a variable effort.\ml{ \citet{ho2015incentivizing} and \citet{Ye2017} instead studied ``effort-responsive tasks''}
In a study where workers are asked to annotate either 5 or 10 items in each HIT, \acite{kazai2011search} shows that better results can be obtained when workers are not overloaded.
Similarly, crowd workers make more errors in counting tasks that include a large number of target objects~\cite{sarma2015surpassing,DasSarma2016}.
Our work intends to systematically evaluate the impact of variable effort on outcome quality by using a task that involves 14 discrete effort levels and requires individual actions for each work unit in the task.

\rev{Other work that focus on task complexity or difficulty has implicitly explored the relationship between task effort and the data quality~\cite{Cai2016,Newell2016,Aipe2018,Yang2016ModelingCrowdsourcing}. For instance, research shows how task ordering can impact the data quality when deploying tasks with varying complexity~\cite{Cai2016}. While these attributes are closely related, task complexity is a different abstraction from the task effort. A task that requires more effort (\eg~annotating an image with 15 faces) is not necessarily more complex than a task that requires less effort (\eg~annotating an image with 2 faces). }

\ms{I suggest that everything from this section up until here be moved into the new related work section about variable effort tasks.}

\subsection{Face Detection Task and Dataset}

For our study, we chose the variable effort task of detecting human faces within an image and drawing bounding boxes around each face. Object detection is one of the most common tasks available on crowdsourcing platforms and is substantially more difficult and time-consuming than simpler tasks like multiple choice questions~\cite{su2012crowdsourcing}.

\begin{figure}[htb]
    \centering
    \begin{subfigure}{0.8\linewidth}
      \includegraphics[width=.45\linewidth]{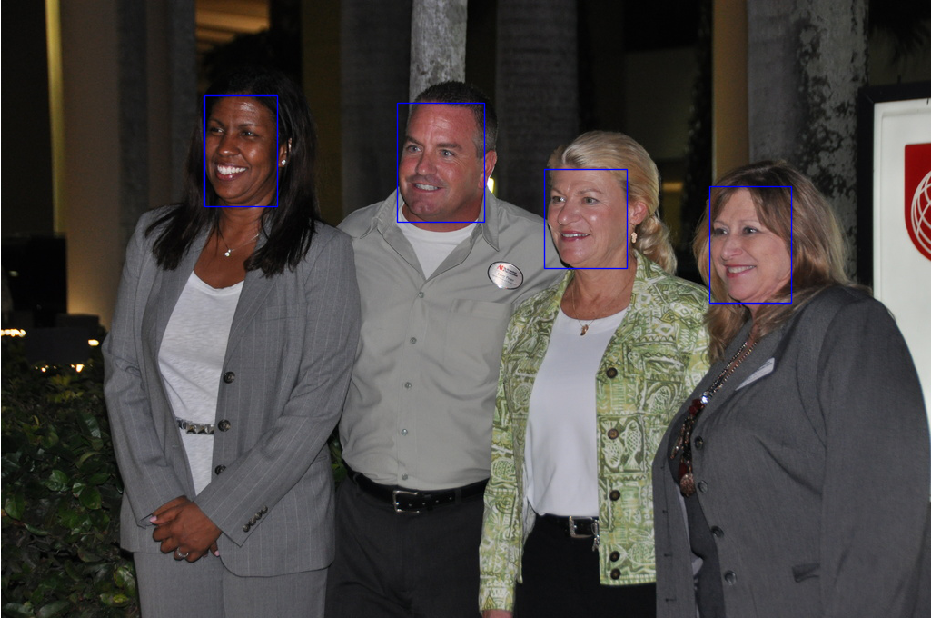}\hfill
      \includegraphics[width=.45\linewidth]{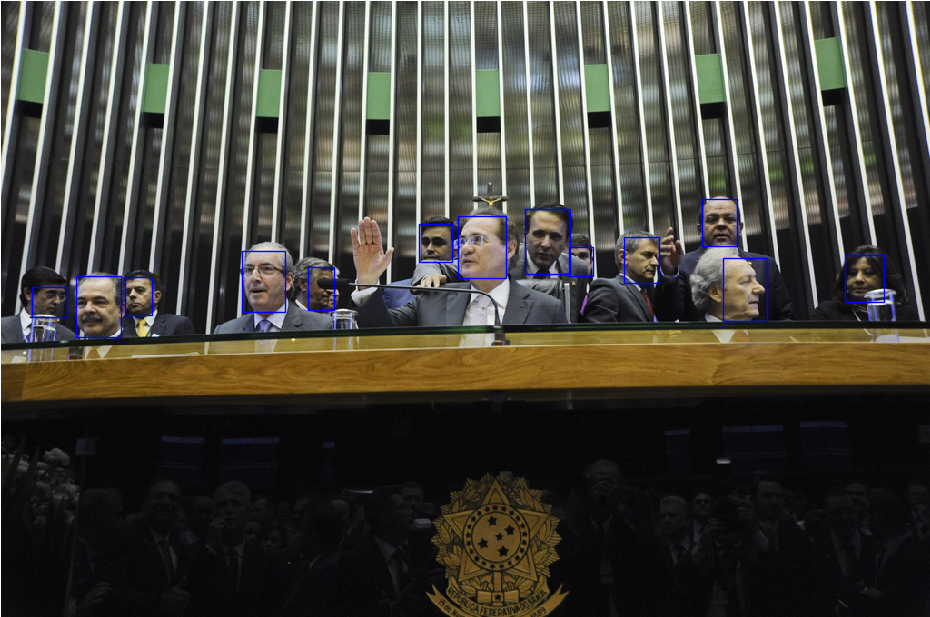}\hfill
      \end{subfigure}
    \caption{Example images with four (left) and ten (right) faces with ground truth data marked using blue rectangles.}
    \label{fig:sample-images}
\end{figure}

We selected 140 images and ground truth data from the Open Images~\cite{OpenImages} dataset.
The number of faces in the images we selected ranged from 1 to 14, with 10 images per face count.
For each subset, we sorted images by ID and picked the first 10 images corresponding to a pseudo-random selection.
Images with potentially ambiguous human faces, such as cartoon characters or statues, were excluded to allow for definitive quality assessments.
\ms{Some of the boxes in Figure~\ref{fig:sample-images} are really hard to see. Can we re-render in a more obvious color? Let's do this after internal submission}
Figure~\ref{fig:sample-images} shows image examples with low and high face counts respectively.

\ml{yellow might be clearer}

Workers completed the face detection task in a standard image annotation interface supporting basic operations such as creating, adjusting and deleting annotations, and zooming. 
We provided consistent base instructions for the annotation task across all experimental conditions, with some condition-specific instructions added where necessary.
The instructions also included three correctly annotated example images~\cite{doroudi2016toward} in each of the conditions.


\subsection{Baseline}

Labeling tasks are typically crowdsourced based on MTurk's pay-per-task pricing model assumes that all task instances are compensated at the same fixed rate. We assume this standard pricing model as our baseline, though we anticipate it may not be optimal for variable effort labeling tasks in which some tasks require much more effort than others. \rev{While a stronger baseline that adjusts payments (\eg~Fair Work~\cite{whiting2019fair}) is desired, we select a more prevalent baseline to ensure the external validity of our results.}


We grouped our image set into two distinct bins based on object count and defined a static payment amount for each bin.
Workers received \$0.16 for completing images with an object count between 1 and 7 and \$0.44 for completing images with an object count between 8 and 14.
This approach assumes an amortized pay of \$0.04 per individual object label.
A more basic alternative would have been to administer a constant pay of \$0.30 per image without binning, but we discarded this design to compare interventions with a more competitive baseline.
We used the standard object detection workflow offered on the Amazon Mechanical Turk platform for our baseline, and no visible gold was administered in this condition.

\subsection{Experimental Setup}

\tm{Danula, in the notebook with the data analysis, there are several cells dedicated to postprocessing the pool of workers to remove obvious spammers I believe. Is that mentioned somewhere in the experimental setup?}
We conducted our crowdsourcing experiments on the Amazon Mechanical Turk (MTurk) platform. All experiments were deployed between 2 PM and 5 PM Pacific Time. \rev{Based on our work time estimates across all conditions, workers on average received an hourly pay of \$10.44, whereas the federal hourly minimum wage in the US is \$7.25.}
Experiments were open to a subset of MTurk workers (a cohort of more than 8000 workers) who had previously qualified for the bounding box task based on a \ms{Reviewers might ask about this quality assurance mechanism. Can we say more about it here? {\bf ML} - this is important} proprietary quality assurance mechanism.
Each experimental condition was available to a unique worker pool created by segmenting the worker subset. Workers were free to complete as few or as many HITs as they liked. \rev{In each condition, images were presented in a random order.} We employed a basic filtering step across all the experimental conditions where we removed obvious spammers by filtering out workers who completed more than 5 HITs and had an average mIoU below 25. We only ended up removing 3 workers across all conditions.

\subsubsection{Measures}

We use \textbf{mean intersection over union (mIoU)} as the primary outcome to compare work quality.
mIoU is a well-established quality metric for object detection tasks~\cite{OpenImages}.
It is computed as the average overall IoU values for all bounding boxes in the ground truth answer key.
For each unmatched ground truth box (false-negative), that box is assigned an IoU of 0. We also report \textbf{task time} as a secondary outcome defined as the time duration from when a worker accepts a task until the task is submitted.
Since we keep the average pay per bounding box constant in all conditions, task payment is not reported as an outcome measure in the paper.

\subsection{Findings and Discussion}

\begin{figure}[htb]
  \centering
  \begin{subfigure}{0.48\linewidth}
      \centering
      \includegraphics[width=1\linewidth]{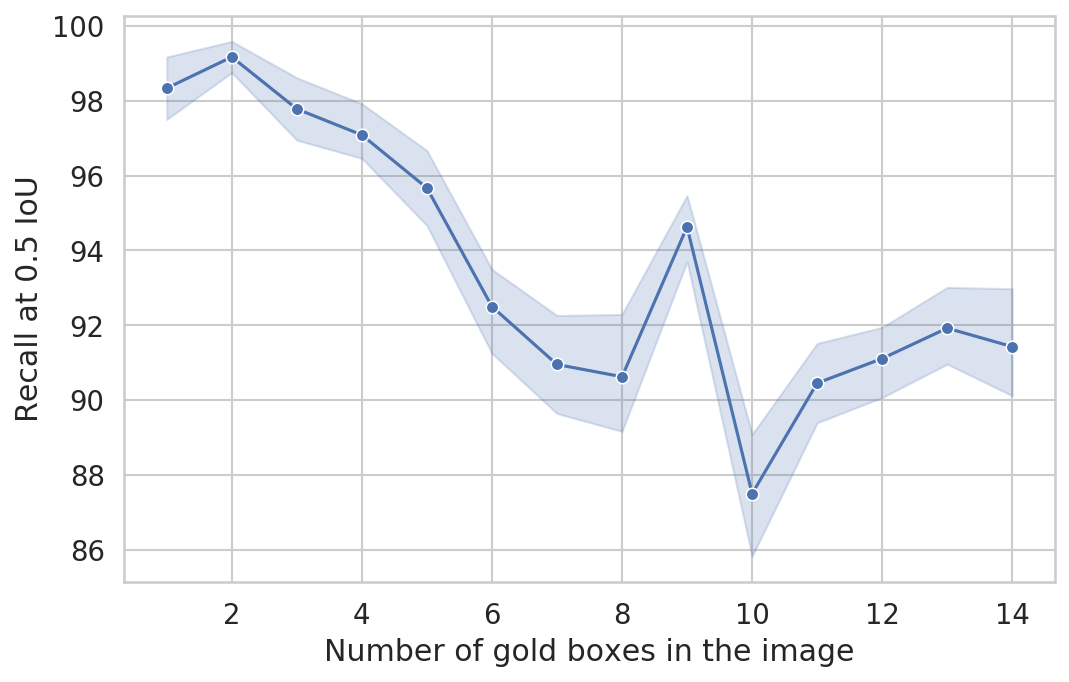}
      \label{fig:misses-effort-0}
  \end{subfigure}
  \begin{subfigure}{0.49\linewidth}
      \centering
      \includegraphics[width=1\linewidth]{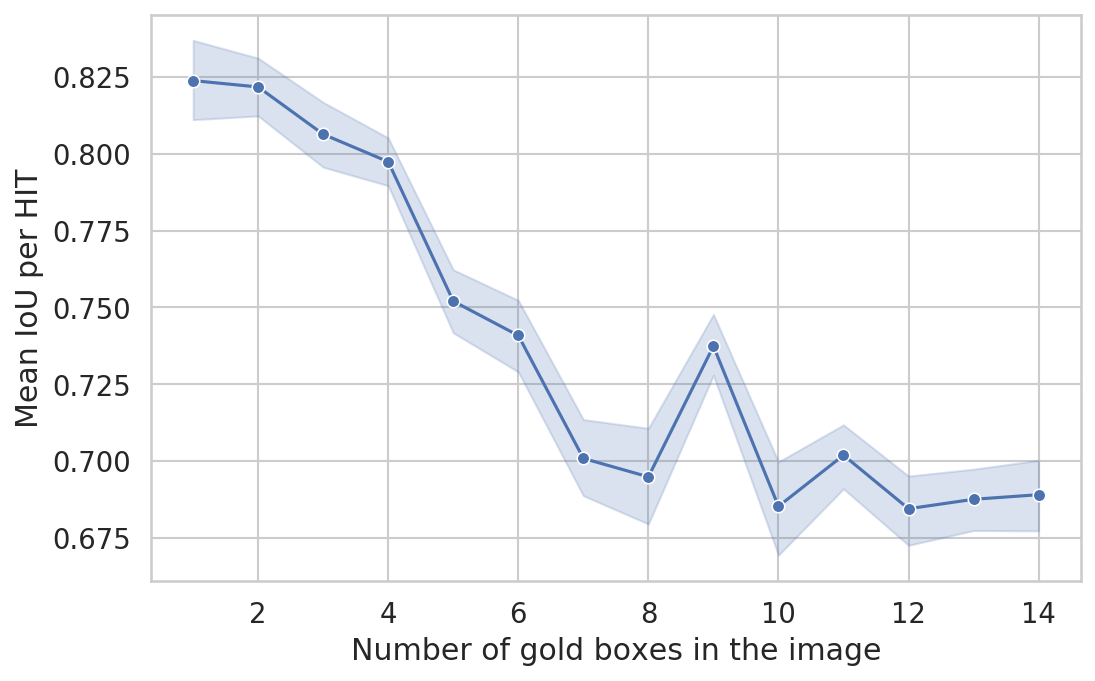}
      \label{fig:mIoU-effort-0}
  \end{subfigure}
  \caption{Variation in output quality against task effort with a naïve baseline of equal pay for each task. \rev{Shaded areas correspond to standard error.}}
  \label{fig:quality-effort-0}
\end{figure}

\rev{We obtained responses from 24 unique workers for the baseline.} Figure~\ref{fig:quality-effort-0} shows that as the number of faces per image increases, annotation quality (as measured by accuracy and recall) declines\ml{explain peak at 9}.\ml{this error analysis is terse and speculative. Can we say concretely what we found?} Data quality reduction in our object detection task can attribute to either workers producing annotations of low-quality or entirely missing particular target objects when there are many target objects present in the image.

As noted above, the baseline assumes fixed-pay for all instances, despite the variable effort required. It could be argued that variation in required effort amortizes over multiple tasks to produce fair pay when workers complete enough work. However, this assumption is questionable. First, prior work shows that requesters often produce poor estimates of average effort per item and tend to underestimate the true effort~\cite{cheng2015measuring}.
This can hinder the administration of fair pay in a systematic manner.
Second, work on crowdsourcing platforms typically follows a power-law distribution where only a few workers complete the majority of work, and the majority of workers abandon a task early~\cite{han2019all}.
Hence, the amortization assumption may only hold true for a small portion of the worker population while the remaining workers receive pay disproportionate to their effort.
Third, even if workers are initially motivated to complete a large number of HITs, drop-out may be encouraged if the first few tasks happen to require high effort.

While high label quality is always desired, the consistency in degradation observed in proportion to the increasing effort is noteworthy. Firstly, uniform labeling quality across all items is desirable, without any consistent biases. Secondly, if we imagine training a detection model on this data, it is particularly important to have accurately annotated images with larger object counts~\cite{shao2019objects365}.
%

\section{Investigating Task Designs for Variable Effort Labeling Tasks}

Section \ref{sec:defining} suggested three key challenges with variable effort tasks: inconsistent worker experience, the potential for high cognitive load, and effective incentive design. In this section, we investigate the potential of various {\em best practice} task designs for crowdsourced annotation to address these challenges.

To investigate the general question of how standard quality improvement methods perform with variable effort annotation tasks.
We picked several crowdsourcing data quality improvement methods that are generalizable and straightforward to implement.
Since appropriate pay for effort is a key driver for quality in crowdsourcing~\cite{kazai2011search}, we chose to include two data collection designs---variable pay and post-task bonus---that aim to calibrate pay according to the required effort on a per-image basis.
Following popular iterative and parallel design paradigms in crowdsourcing, we included two other designs---task decomposition and iterative improvement---that aim to standardize the required effort in each task unit.
Finally, we add the visible gold design, which uses gold standard questions for testing and training.
For each of the data collection designs outlined below, we collected three responses per image.

\subsection{Variable Pay}

For better per-item calibration of payment, a more sophisticated data collection design may aim to estimate the effort for each item in advance and adjust the payment on a per-item basis accordingly.
These estimates can be produced either manually (e.g., via upstream crowdsourcing workflows~\cite{Bernstein2010Soylent}) or automatically (e.g., via pre-built object detection models~\cite{borji2019salient}).

However, this data collection design introduces other challenges.
First, a priori estimation of effort is a non-trivial task and can be costly when manual workflows are required.
Second, since HIT payment is one of the parameters used by Amazon Mechanical Turk to group HITs in the platform, tasks with different effort levels are advertised separately, allowing workers to selectively focus on tasks with high pay and ignore low paying tasks.

We instantiated this data collection design in our study by setting the payment amount for each image proportional to the exact object count as defined in the available ground truth data.
In particular, we pre-calculated pay per image by multiplying the true object count with the base pay of \$0.04 per object (e.g., \$0.04 for images with a single object and \$0.56 for images with 14 objects).

\subsection{Post-task Bonus}

An alternative to a priori effort estimation is to decide an appropriate payment amount \textit{after} a task has been completed~\cite{yin2015bonus}.
This can be accomplished by setting up a HIT with a flat base payment and advertising a post-task bonus to compensate workers for any work completed beyond the base payment.

There are two main challenges to this data collection design pertaining to the trust relationship between workers and requesters.
On the one hand, workers need to trust a requester to deliver on their promise of administering a post-hoc bonus and to choose the bonus amount fairly.
On the other hand, requesters rely on good-faith execution of the task (e.g., the number of objects labeled or time spent) to produce accurate estimates of effort and fair bonus amounts.
To this end, labels can be verified in a secondary process, e.g., manual verification through other workers, incurring additional cost for the requester.

We implemented this data collection design leveraging ground truth information available in our dataset.
In particular, we offered workers a total payment of \$0.04 for each object labeled correctly as per the ground truth data.
The flat base payment was set to \$0.04 for all images difference between flat base payment, and total payment was administered as the post-task bonus.

\subsection{Task Decomposition}

The previous two data collection designs aim at adjusting payment to variable effort on a per-item basis.
An alternative approach is to decompose tasks into fixed-size units with constant effort and to administer a constant pay amount per task unit.

Prior work has shown that task decomposition not only facilitates fair payment, but also aids workers in producing high-quality answers by managing cognitive load~\cite{sarma2015surpassing}.
However, the process of decomposing tasks into smaller chunks of equal size is non-trivial, and prior work suggests that some tasks depend on the context, which may not be preserved during decomposition~\cite{Tong2018SLADE:Crowdsourcing}. Getting multiple individuals to attempt smaller portions of the same task can also help elevate the output quality~\cite{sarma2015surpassing,kittur2011crowdforge,teevan2016supporting}.

We implemented this data collection design using a two-step workflow.
First, we identified all target objects in a given image.
Second, we created sub-tasks where workers were asked to create bounding boxes with a pre-defined set of target objects indicated through point markers.
In our experiment, we created HITs with a maximum of 3 target objects and a static pay of \$0.08 per HIT corresponding to 2 target objects per image on average.
We implemented two variants for identifying target objects in the initial step:
\begin{itemize}
    \item \textbf{Oracle:} In this variant, target objects were identified using the available ground truth data.
    \item \textbf{Manual:} Since ground truth data is generally not available in practice, we implemented a second variant in which target objects were identified manually by workers through a separate upstream point annotation task.
\end{itemize}

In addition to the above variants, for scalability, task decomposition can also be achieved through automated estimation methods. For example, as the decomposition step does not require high accuracy, we can use a generic automated object detection model~\cite{borji2019salient} to generate the object markers and decompose tasks.

\subsection{Iterative Improvement}

We also included a task design informed by iterative crowdsourcing workflows~\cite{goto2016understanding} where several workers contribute to the same task while each worker can see the results from the previous worker.
Prior work shows that iteration can increase the response quality in specific tasks like writing~\cite{little2010exploring}, brainstorming~\cite{little2010exploring}, and translation~\cite{Ambati2012Collaborative}.
Quality increase typically attributes to corrective actions taken by subsequent workers.
In addition, as our task involves variable effort, we also leverage an iterative workflow to regulate the amount of work that each worker needs to complete in a single iteration.

In our experimental design, multiple workers iteratively annotate the same image. In each iteration, we ask workers to either annotate a maximum of N=3 new objects, adjust existing work or mark the task as completed indicating that there is nothing left to annotate. Each iteration was deployed as a single HIT, and we set a static pay of \$0.08 per HIT corresponding to an average number of two target objects per image.

\subsection{Visible Gold Questions}


Finally, we also explored a scenario where gold standard answers are available for a small subset of items in the dataset.
These items with known answers are injected into workers' task queues to assess annotation quality on an intermittent basis.
In addition to assessing quality, gold standard answers can be used to provide near real-time feedback to workers for each object they labeled (or failed to label), e.g., for each face in an image.
We refer to this feedback mechanism as ``visible gold''. Prior works~\cite{le2010ensuring,gadiraju2015training,doroudi2016toward} have investigated the use of visible gold for simple tasks with binary outcomes or multiple choices and focused on presenting visible gold questions upfront or with a static gold-to-task ratio.

\begin{figure}[htb]
  \centering
  \includegraphics[width=0.85\linewidth]{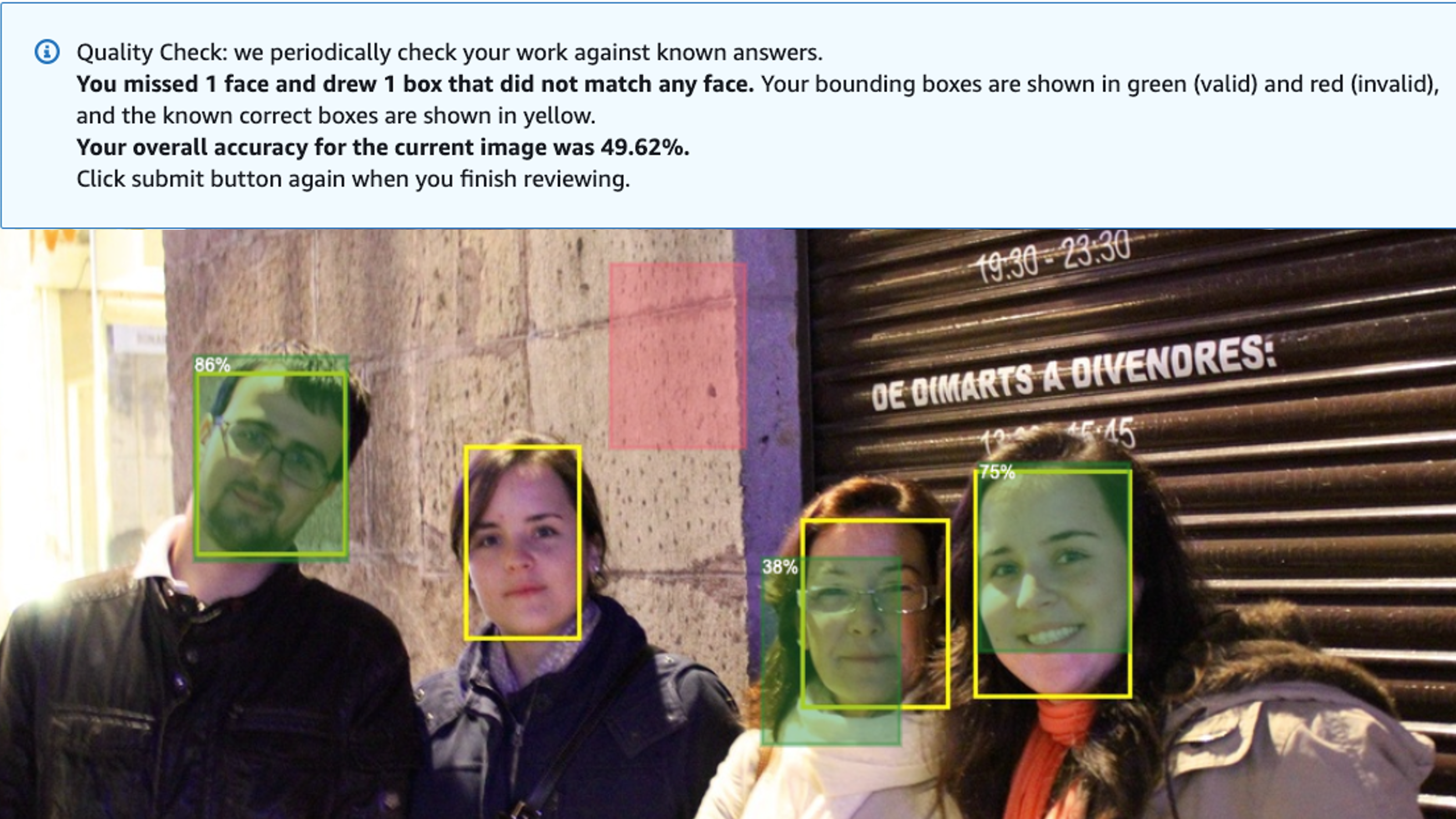}
    \caption{Visible Gold Feedback Interface
    }
    \label{fig:visible-gold-interface}
\end{figure}

Figure~\ref{fig:visible-gold-interface} shows our interface to provide feedback to workers using visible gold for the variable effort annotation task used in our study.
Workers encounter the interface immediately after submitting their answer to a visible gold question.
The feedback provided to workers included the number of objects missed (i.e., false negatives), the number of bounding boxes drawn by the worker that did not match any objects in the answer key (i.e., false positives), the accuracy for each bounding box correctly matching an object (i.e., for each true positive) and the average accuracy across all bounding boxes in the answer key (using 0\% accuracy for false negatives).
Accuracy for individual bounding boxes was calculated as ``intersection over union'' (IoU), i.e., the ratio between the area of overlap and the area of union of the worker annotation and the ground truth bounding box.
IoU is a standard accuracy measure in object detection tasks.
Finally, all ground truth bounding boxes were displayed to workers along with their own annotations as part of the feedback interface.
While gold standard annotations were available for all images in our study, the interface was designed to dynamically decide whether a particular question should be a visible gold for a given worker based on the experimental condition.

As workers get immediate performance metrics through visible golds, they are motivated to produce high-quality work even when the task involves more effort.
In addition, through the detailed feedback mechanism, visible golds can provide increased task clarity and help workers understand and improve their task performance.
We expect that these characteristics will help set accurate expectations and motivate workers to carefully attempt tasks that involve an increased effort.

\section{Evaluation}


We conducted our crowdsourcing experiments on the Amazon Mechanical Turk (MTurk) platform and used a consistent experimental setup as described in Section 3.4.
On average, workers received a payment of \$0.04 per bounding box (e.g., a worker receives a payment of \$0.4 for a HIT that includes an image with 10 bounding boxes).

\subsection{Results}

Table \ref{tab:summary-results-1} presents a summary of results, including the mean and standard error for mIoU values and average task time for each condition.
\rev{
In each condition, mIoU values follow a non-normal distribution. A Mann Whitney test with Bonferroni correction for multiple comparisons shows that mIoU values are significantly lower in iterative improvement ($M=55.6$), $U=499197.0$, $p<0.001$, post-task bonus ($M=59.5$), $U=221215.0$, $p<0.001$, task decomposition oracle ($M=67.5$), $U=286194.0$, $p<0.001$, variable pay ($M=69.1$), $U=289741$, $p<0.001$ and not significantly different to task decomposition manual ($M=71.7$), $U=330296.0$, $p=0.21$ compared to the baseline ($M=73.7$). However, mIoU values in visible gold regular ($M=75.5$), $U=397034.0$, $p<0.01$ are significantly higher compared to the baseline.
}

\ml{can we reduce precision to mIoU=XX.X, SE 0.XX, and time=XXX.X?}

\begin{table}[h]
  \caption{Mean IoU across conditions.}
  \label{tab:summary-results-1}
  \begin{tabular}{lrrr}
    \toprule
    &&& Average Task \\
    Condition & Mean (mIoU) & SE (mIoU) & Time (sec)\\
    \midrule
    Baseline  &	73.7 & 0.46 & 182.6\\
    \midrule
    Iterative Improvement & 55.6 & 0.80 & {\bf 164.4}\\  
    Post-task Bonus on work load & 59.5 & 1.08 & 243.4\\
    Task Decomposition Manual & 71.7 & 0.74 & 613.8\\
    Task Decomposition Oracle & 67.5 & 0.90 & 557.7\\
    Variable Pay on work load & 69.1 & 0.79 & 227.2\\
   Visible Gold - Regular & \textbf{75.5} &	0.58 & 168.4 \\
   \bottomrule
\end{tabular}
\end{table}

Figure~\ref{fig:time} shows how task time per bounding box varies according to the number of ground truth target objects available in the image.

\begin{figure}[h]
    
\end{figure}

\subsubsection{Impact of Task effort}

\begin{figure}[htb]
  \centering
  \begin{subfigure}{0.29\linewidth}
      \centering
      \includegraphics[width=1\linewidth]{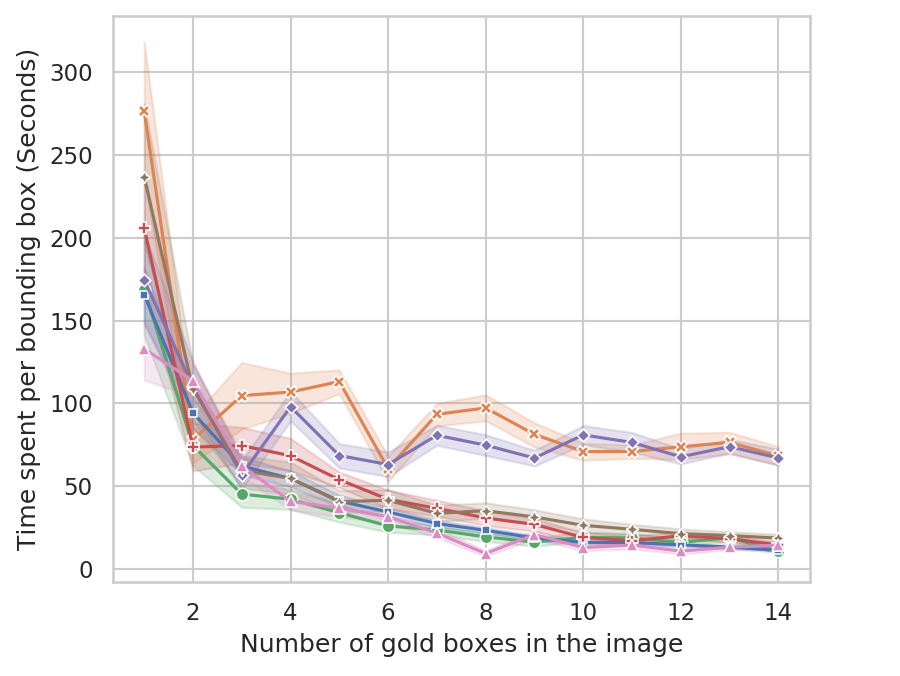}
      \caption{}
      \label{fig:time}
  \end{subfigure}
  \begin{subfigure}{0.27\linewidth}
      \centering
      \includegraphics[width=1\linewidth]{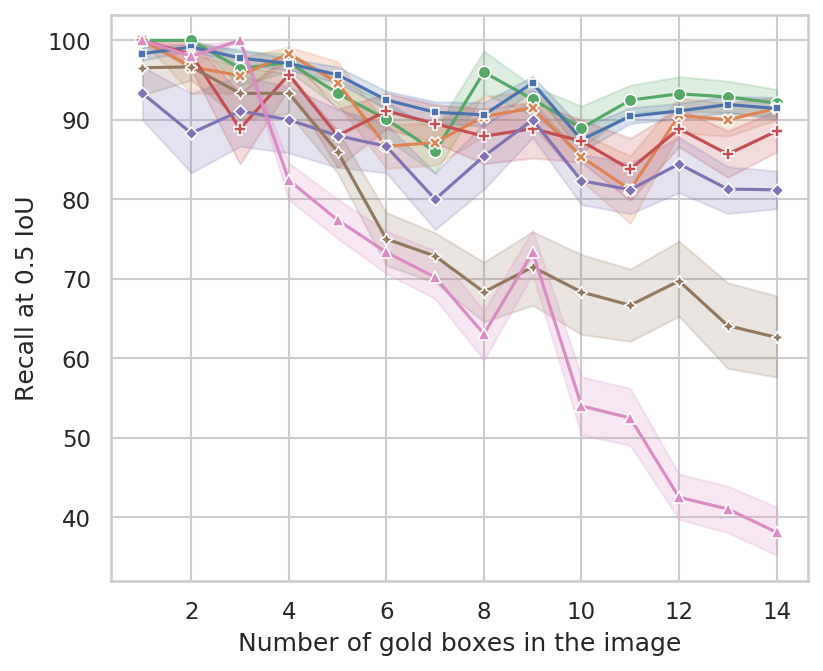}
      \caption{}
      \label{fig:misses-effort-1}
  \end{subfigure}
  \begin{subfigure}{0.42\linewidth}
      \centering
      \includegraphics[width=1\linewidth]{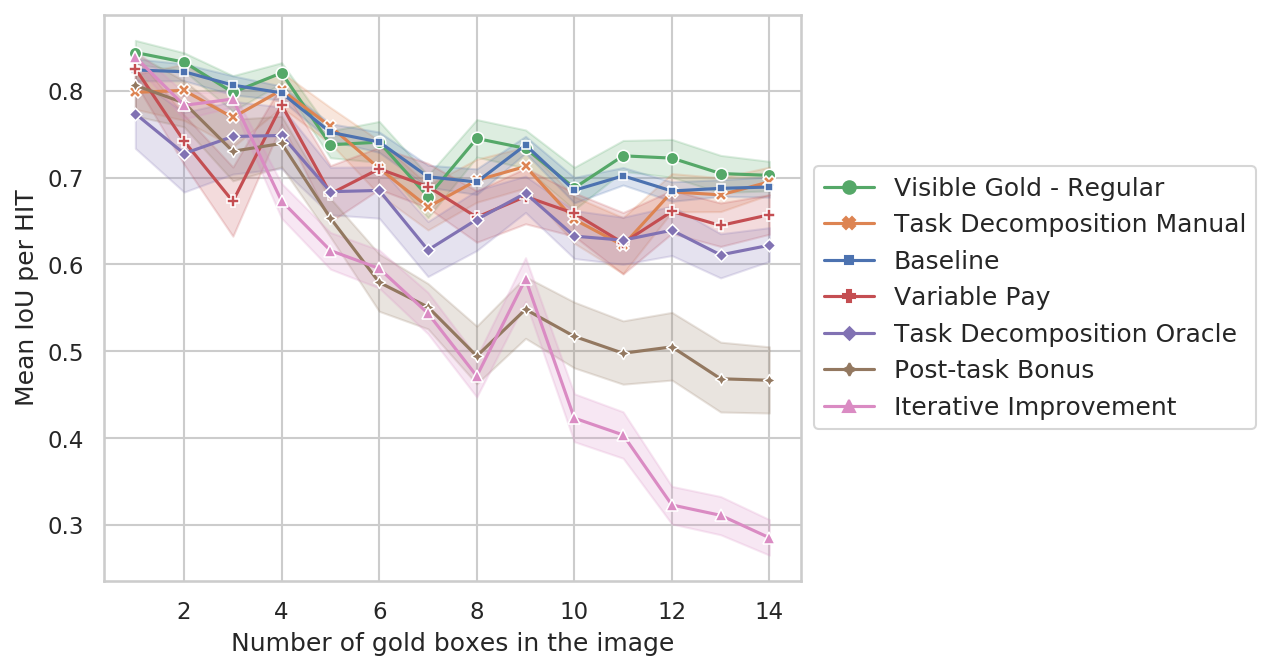}
      \caption{}
      \label{fig:mIoU-effort-1}
  \end{subfigure}
  \caption{Variation in task time and output quality against task effort across conditions. \rev{Shaded areas correspond to standard error.}}
  \label{fig:quality-effort-1}
\end{figure}

Complex crowdsourcing tasks also include variable efforts within HITs.
In object annotation, the number of bounding boxes that needs to be annotated in each image can range from 1 to many.
In Figure~\ref{fig:quality-effort-1}, we examine how the output quality varies when the task effort increases.
We see that our improved visible gold method consistently outperforms other methods in terms of mIoU (Figure~\ref{fig:mIoU-effort-1}) and recall at mIoU>0.5 (Figure~\ref{fig:misses-effort-1}).

\begin{figure}[htb]
  \centering
  \includegraphics[width=0.55\linewidth]{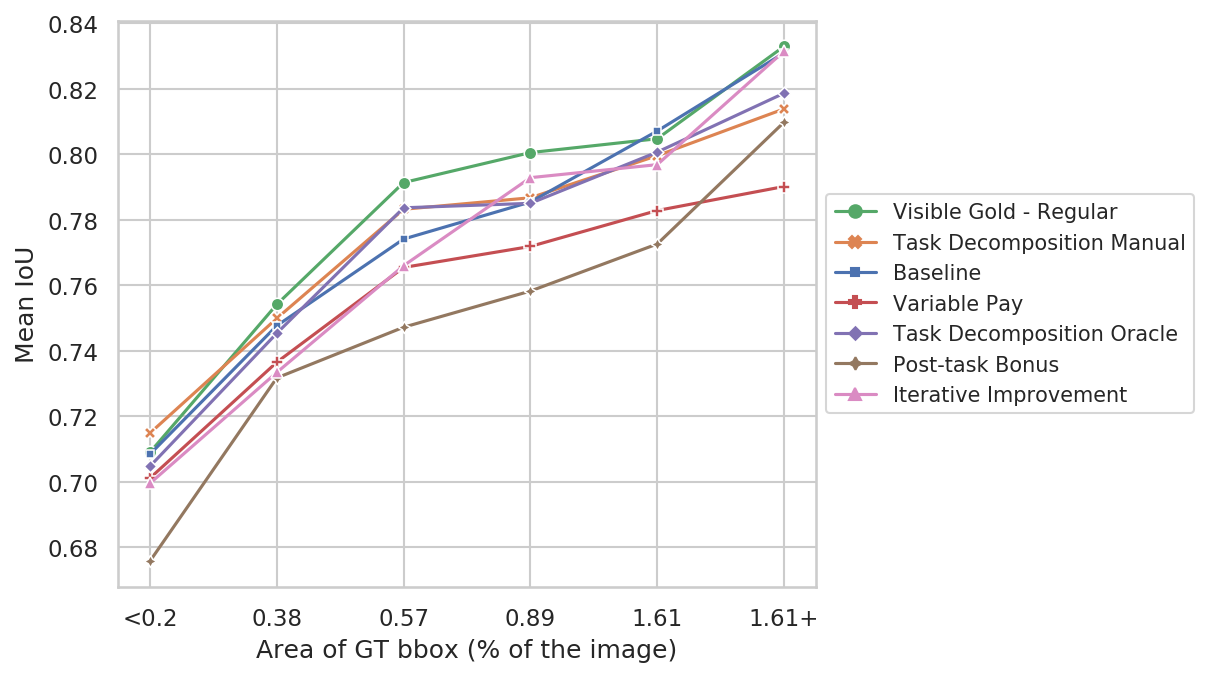}
  \caption{Variation in mIoU against object size across conditions.}
  \label{fig:mIoU-target-size-1}
\end{figure}

Figure~\ref{fig:mIoU-target-size-1} shows how mIoU varies depending on the size of each ground truth bounding box.
The worker output quality is relatively low for smaller objects.
While this trend is visible across all the methods, the improved visible gold method performs well above the other methods regardless of the target object size.

\subsubsection{Task Completion}

\begin{figure}[h]
    \centering
    \includegraphics[width=0.8\textwidth]{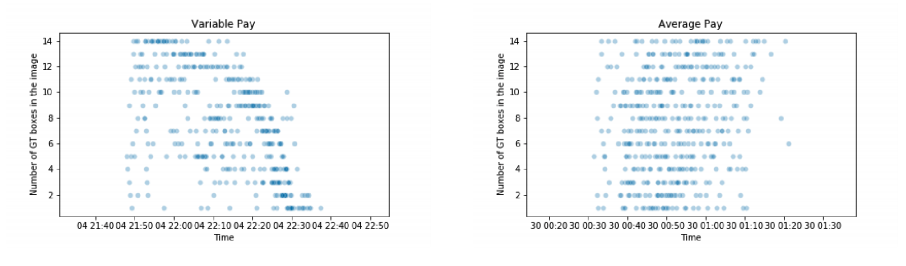}
    \caption{Variation in task completion between the baseline (average pay) and the variable pay condition.}
    \label{fig:task-completion}
\end{figure}

Task completion patterns based on the task submit time for variable pay and baseline conditions are given in Figure~\ref{fig:task-completion}.

\danh{Change Average Pay -> Baseline}

\subsection{Analysis of Findings}

Based on the results presented in Table~\ref{tab:summary-results-1}, visible gold method results in the highest mIoU across all the attempted quality improvement methods. All the other quality improvement methods fail to surpass the baseline when considering the mIoU value.

To produce high-quality annotations, we also want crowd workers to carefully complete the task by spending ample time. As seen in Figure~\ref{fig:time}, time spent on a work unit or a single object annotation diminishes when the number of objects in the image increases. On the contrary, task decomposition conditions that appear as outliers in Figure~\ref{fig:time} can ensure consistent work time on each unit when compared to other conditions. However, we did not obtain the desired quality improvement. As given in Table~\ref{tab:summary-results-1}, mIoU values from task decomposition conditions are lower than the baseline condition.

As shown in Figure~\ref{fig:quality-effort-1}, output quality measured in mIoU and the percentage of annotated ground truth boxes drops as the number of target objects in images increases. While this general trend is present in all experiment conditions, Visible Gold, Baseline, and Task Decomposition (Manual and Oracle) perform relatively better compared to other methods. Output quality drastically drops with effort in iterative improvement and post-task bonus conditions. For post-task bonus, this is mainly due to the low task base-price. For instance, when a worker accurately labels an image with 12 target objects, they receive a base pay of \$0.08 and a bonus of \$0.40. Although we use a reputed requester account with a 99\% approval rate, we can argue that workers are still unwilling to commit to a task with low specified payment and do additional work without a payment guarantee. The reason behind the observation in iterative improvement condition is not explicit. Prior work also notes that data quality can be reduced when subsequent workers are led down the wrong path in an iterative workflow of tasks with high difficulty~\cite{little2010exploring}. Another possible cause can be workers failing to understand the task/instructions fully and confusing it with standard verification jobs, and prematurely marking the task as completed.

As a general trend, mIoU in our object detection task decreases when the target object is smaller.
There are two possible causes for this observation.
First, workers could completely miss smaller target objects during the annotation process when there are many target objects present in the image (as seen in Figure~\ref{fig:quality-effort-1}).
Second, when the object is smaller, the impact on error is also higher (e.g., the error caused by missing the margin by a single pixel is higher for relatively smaller target objects).

In contrast to other conditions, workers could pick tasks with specific object counts under variable pay condition. In Figure~\ref{fig:task-completion}, we observe that workers prioritized tasks with higher effort and higher pay. 

\section{Improving Visible Gold}

\ml{we deep dive on visible gold because both 1) it performed best; and 2) it is perhaps more general to benefit non-variable effort tasks as well, and more general solutions are preferrable}

Our initial evaluation shows how visible gold method can result in annotations with higher quality in object detection task when compared to other quality improvement methods. Also, visible gold method that relies on gold standard questions is potentially more generalizable across many task types as opposed to other methods like task decomposition and iterative improvement. The applicability of visible gold is also not limited to tasks that involve a variable effort. These factors led us to further investigate visible gold as a promising generic quality improvement method for crowdsourcing.

In this section, we detail how we refined our visible gold method. First, we explore different visible gold issuing patterns. Second, we evaluate how bonuses and warnings work as consequences when using visible gold. Finally, we present an improved visible gold method that incorporates tier-based consequences and dynamic visible gold issuing.

\subsection{Visible Gold Issuing Pattern} 
We evaluate three ways of issuing visible gold questions and obtained five responses per image in each condition.

\begin{itemize}
    \item \textit{Upfront}: Workers complete a fixed number of visible gold questions at the beginning. This condition is comparable to the explicit training in previous work by \acite{gadiraju2015training}. Upfront condition is straightforward to implement and can be considered as a variant of a qualification test.
    \item \textit{Regular}: Workers encounter a visible gold regularly with a visible gold to task ratio of 1:4 similar to the prior work by ~\acite{le2010ensuring}. We also included an offset such that workers are not able to anticipate a visible gold by following the issuing pattern.
    \item \textit{Fib+Regular}: We propose a new strategy that combines the characteristics of Upfront and Regular conditions. We seek to present more visible gold questions at the beginning such that we can test the workers reliably while providing ample training examples. But as workers continue, we want to test less frequently. We achieve this by following the Fibonacci sequence. Under this condition, up to 50 questions, we follow the Fibonacci sequence (\ie~ 1,2,3,5,8,13..) to issue visible golds and then falls to a more infrequent regular visible gold to task ratio of 1:19.\ml{explain rationale for fibonnaci: more testing early and less over time, and why this seems good}
\end{itemize}

\subsection{Bonus vs. Warning as a consequence} 

For Upfront, Regular and Fib+Regular conditions, we used warning as the consequence where we warned workers that they would not be able to attempt future tasks if their outcome measured via quality checks (i.e., visible gold questions) does not meet the expected quality standard. However, we did not block any workers during the task or remove any contributions from our analysis. To examine if incentivizing workers to produce high-quality annotations works better than the warning, we deployed the following additional condition.

\begin{itemize}
    \item \textit{Regular Bonus}: Similar to the `Regular' condition, workers encounter a visible gold regularly with a visible gold to task ratio of 1:4. For the consequence, instead of the warning, workers receive a bonus payment of \$0.08 (face count less than 8) or \$0.22 (face count greater than 7) per image if they maintain an accuracy above a pre-specified threshold. Bonus amounts and thresholds were specified in the task description.
\end{itemize}

\subsection{Dynamic visible gold and tier-based consequences}

Based on the literature and results obtained in our first round of experiments detailed above, we further improved our visible gold mechanism by dynamically adjusting the visible gold issuing pattern and by adding a performance metric display element.

\begin{figure}[htb]
  \centering
  \includegraphics[width=0.9\linewidth]{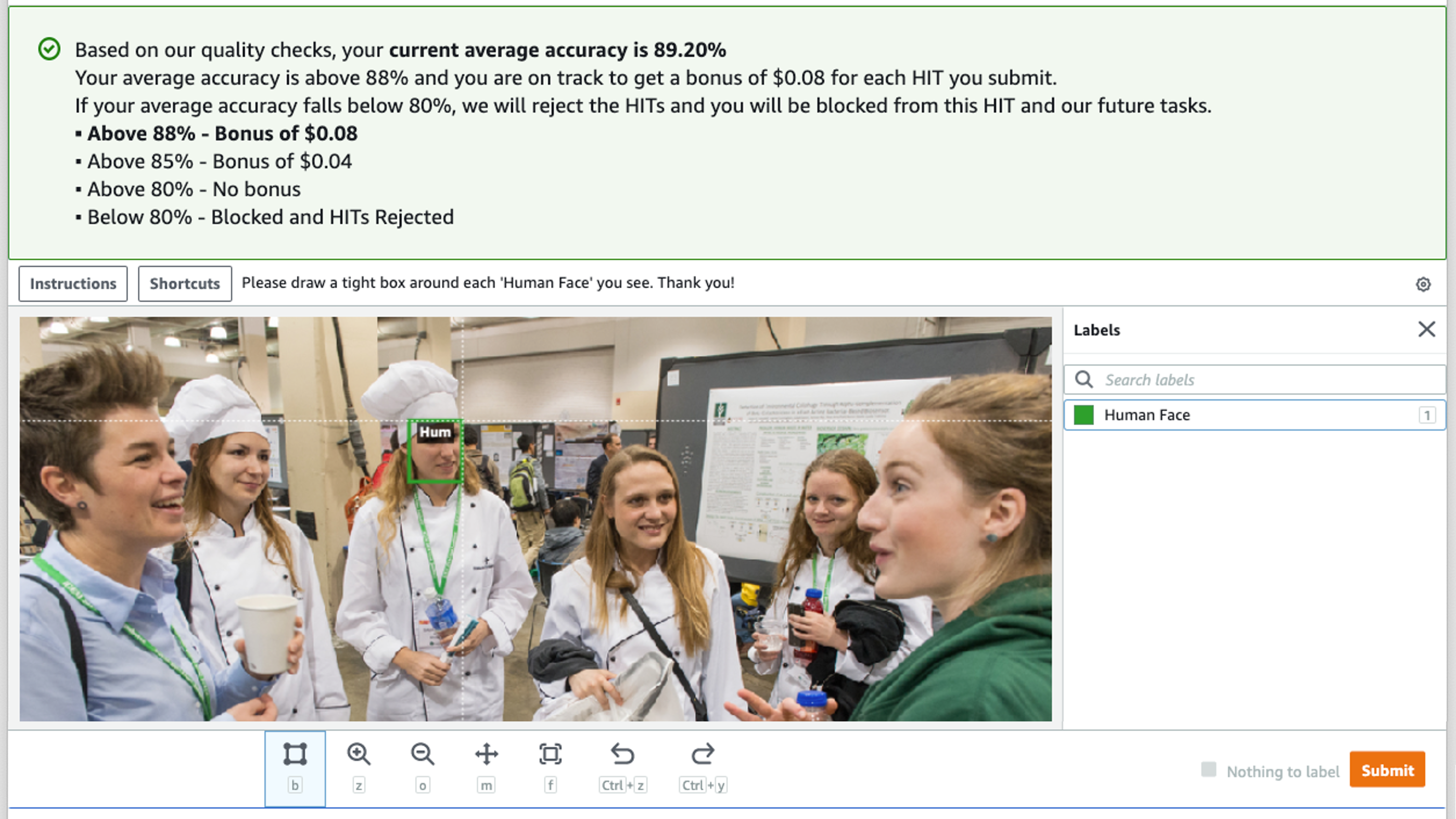}
    \caption{Improved bounding box task interface}
    \label{fig:task-interface}
\end{figure}

During our initial experiments, a handful of workers ignored the training and guidance provided by visible gold questions and continued to produce low-quality work.
To counter this, we altered the Fib+Regular visible gold issuing pattern by adding bonusing and blocking conditions with pre-defined quality thresholds.
Bonus thresholds $T_{BONUS-A}$ and $T_{BONUS-B}$ determined whether a worker qualified for a bonus payment.
Blocking threshold $T_{MIN}$ was the minimum mIoU that a worker needed to achieve to pass a given visible gold.
When a worker completed a visible gold with their mIoU for the current image falling below $T_{MIN}$, we overrode the standard visible gold pattern and issued another visible gold as the next HIT.
\ms{Which visible gold pattern was used for conditions "Visible Gold - Improved" and "Visible Gold - Regular Bonus"? Upfront, Regular or Fib+Regular?}
\ml{Improved is not a descriptive name for what was done}
We continued to issue visible golds until either the worker passed a visible gold or they met the blocking condition by failing three consecutive visible golds with an overall average accuracy below the blocking threshold $T_{MIN}$.

To increase transparency, we added a dedicated performance metric banner at the top of the task interface as seen in Figure~\ref{fig:task-interface}.
A worker could see their current average accuracy and the relevant quality tier in a simplified manner.
The content of the banner was updated as workers completed visible golds.

\ms{I'm confused about the various conditions at this point. "Baseline", "Upfront" and "Regular" are clear. "Fib+Regular" is the same as "Fixed", right? I'm not clear on which visible gold pattern was used as the basis for "Regular Bonus" and "Improved". I also assume that we don't use any visible gold in any of the "Other Conditions", but this isn't 100\% clear yet.}
We ran additional experiments with the improved visible gold interface.
We collected five responses per image and kept all other parameters regarding the experiment consistent with previous deployments detailed in Crowdsourcing Experiments (Section 3.4).

\section{Evaluation II}

\subsection{Results}

A summary of results for visible gold conditions, including the mean and standard error for mIoU values and average task time is given in Table~\ref{tab:summary-results-2}.

\begin{table}[htb]
  \caption{Mean IoU across conditions.}
  \label{tab:summary-results-2}
  \begin{tabular}{lrrr}
    \toprule
    &&& Average Task \\
    Condition & Mean (mIoU) & SE (mIoU) & Time (sec)\\
    \midrule
    Baseline  &	73.7 & 0.46 & 182.6\\
    Visible Gold - Regular & 75.5 &	0.58 & 168.4 \\
    \midrule
    Visible Gold - Upfront & 75.5 & 0.53 & {\bf 165.6} \\
    Visible Gold - Fib+Regular	& 75.7 & 0.58 & 177.0\\
    Visible Gold - Regular Bonus & 74.7 & 0.59 & 193.9\\
    Visible Gold - Improved & \textbf{79.3} & 0.41 & 168.0\\
  \bottomrule
\end{tabular}
\end{table}

\begin{figure}[h]
    \centering
    \includegraphics[width=0.8\linewidth]{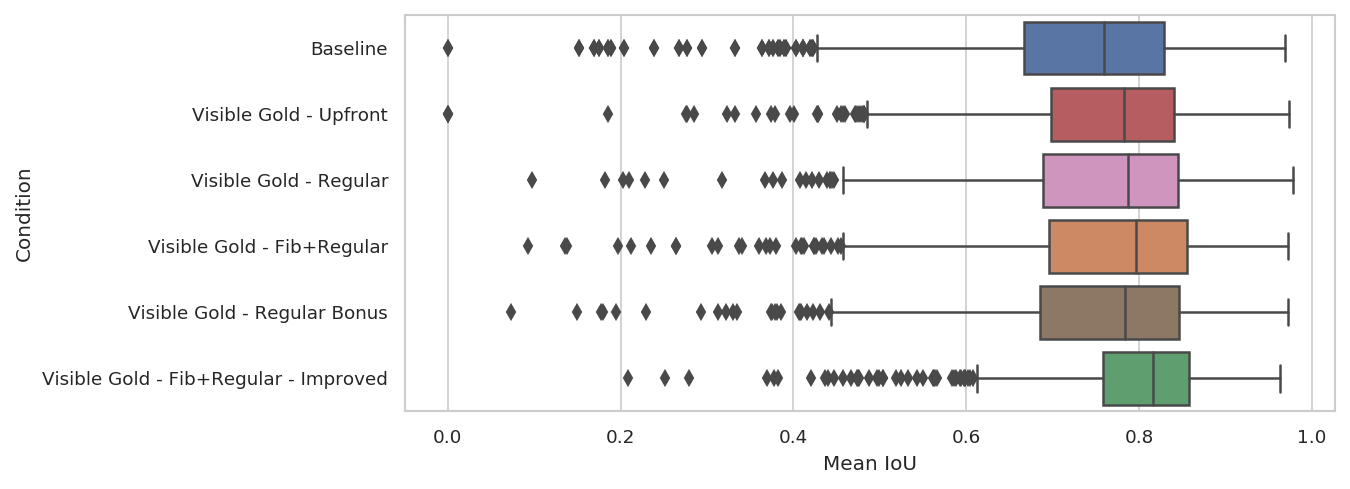}
    \caption{Distribution of mIoU for each submission across conditions}
    \label{fig:gold-histogram}
\end{figure}
\ml{could remove ``Visible Gold'' from each label to increase plot size}

\subsubsection{Visible Gold Execution}

From our initial experiments to identify the suitable visible gold execution strategy, all strategies produced better outcomes when compared to the baseline.
\tm{Are these stats corrected for multiple comparisons?} \rev{Obtained mIoU values in each condition follow a non-normal distribution.} A Mann Whitney test with Bonferroni correction for multiple comparisons shows that mIoU visible-gold-upfront ($M=75.447$), $U=483319$, $p<0.01$, visible-gold-regular ($M=75.48$), $U=397034$, $p<0.01$, and visible-gold-fib+regular ($M=75.650$), $U=451168$, $p<0.001$ have significantly higher mIoU values compared to the baseline ($M=73.739$).

\begin{figure}[h]
  \centering
  \begin{subfigure}{0.49\linewidth}
      \centering
      \includegraphics[width=.9\linewidth]{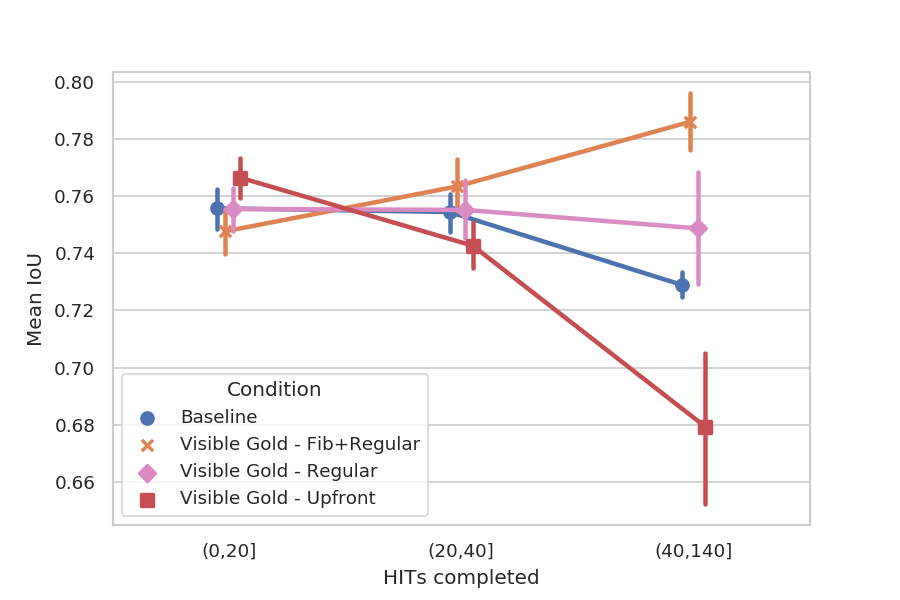}
      \caption{Variation in mIOU against HIT completion.\newline \rev{Error bars show standard error.}}
      \label{fig:mIoU-hit-order}
  \end{subfigure}
  \begin{subfigure}{0.49\linewidth}
      \centering
      \includegraphics[width=.9\linewidth]{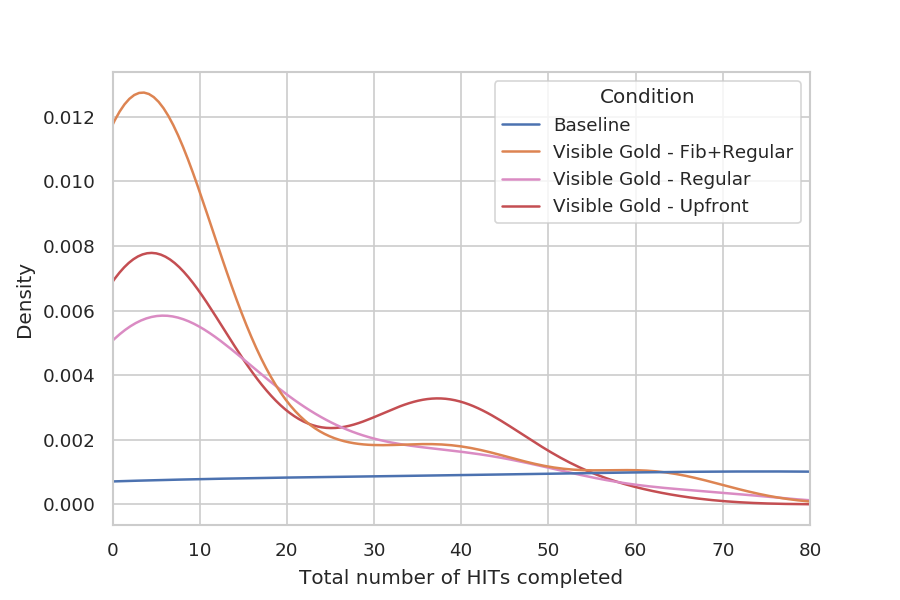}
      \caption{Distribution of the total number of HITs completed by each worker}
      \label{fig:workers-hitscompleted}
  \end{subfigure}
  \caption{\rev{HIT completion across conditions}}
\end{figure}

In order to identify the most suitable visible gold execution strategy, we separated answers into three bins based on the hit completion order and plotted the mIoU metric in Figure~\ref{fig:mIoU-hit-order}. In Figure~\ref{fig:workers-hitscompleted}, we also show the variation in the distribution of the total number of tasks completed by each worker.

\ml{where is ``improved''?}


\ml{question I anticipate: did individual performance improve over time, and how was 7\% improvement distributed? Did this design just get workers to stay longer, and those who did learned to do better?}

\subsubsection{Visible Gold Improved}

A Mann Whitney test indicated that mean mIoU in improved visible gold condition ($M=79.301$) is significantly higher than mean mIoU in baseline ($M=73.739$), $U=393812$, $p<0.001$.
Results in improved condition were also significantly higher than visible-gold-fib+regular condition ($M=75.650$), $U=172731$, $p<0.001$, which provided the best outcome in the first round of experiments.



\subsubsection{Bonus vs. Warning}

We compare between using warnings and bonuses as a consequence for failing visible gold questions.
Our results show that there is no significant difference between bonus ($M=74.670$) and warning ($M=75.483$) conditions, $U=145062$, $p=0.26$.

\subsubsection{Impact of Task effort}

\begin{figure}[htb]
  \centering
  \begin{subfigure}{0.395\linewidth}
      \centering
      \includegraphics[width=1\linewidth]{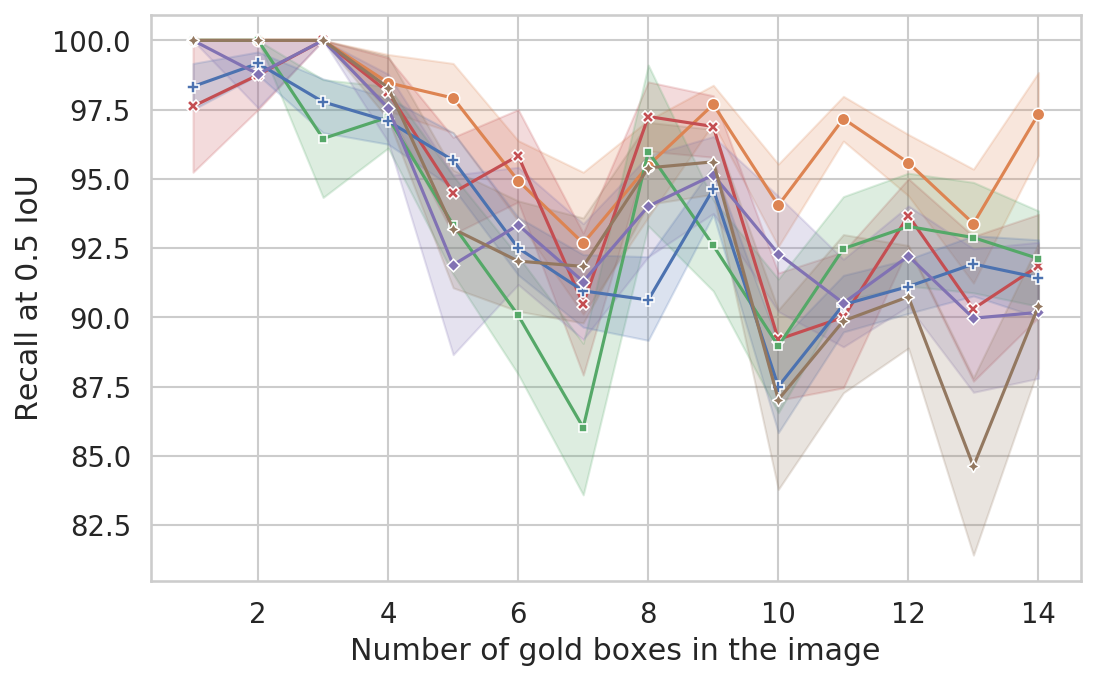}
      \caption{}
      \label{fig:misses-workload}
  \end{subfigure}
  \begin{subfigure}{0.595\linewidth}
      \centering
      \includegraphics[width=1\linewidth]{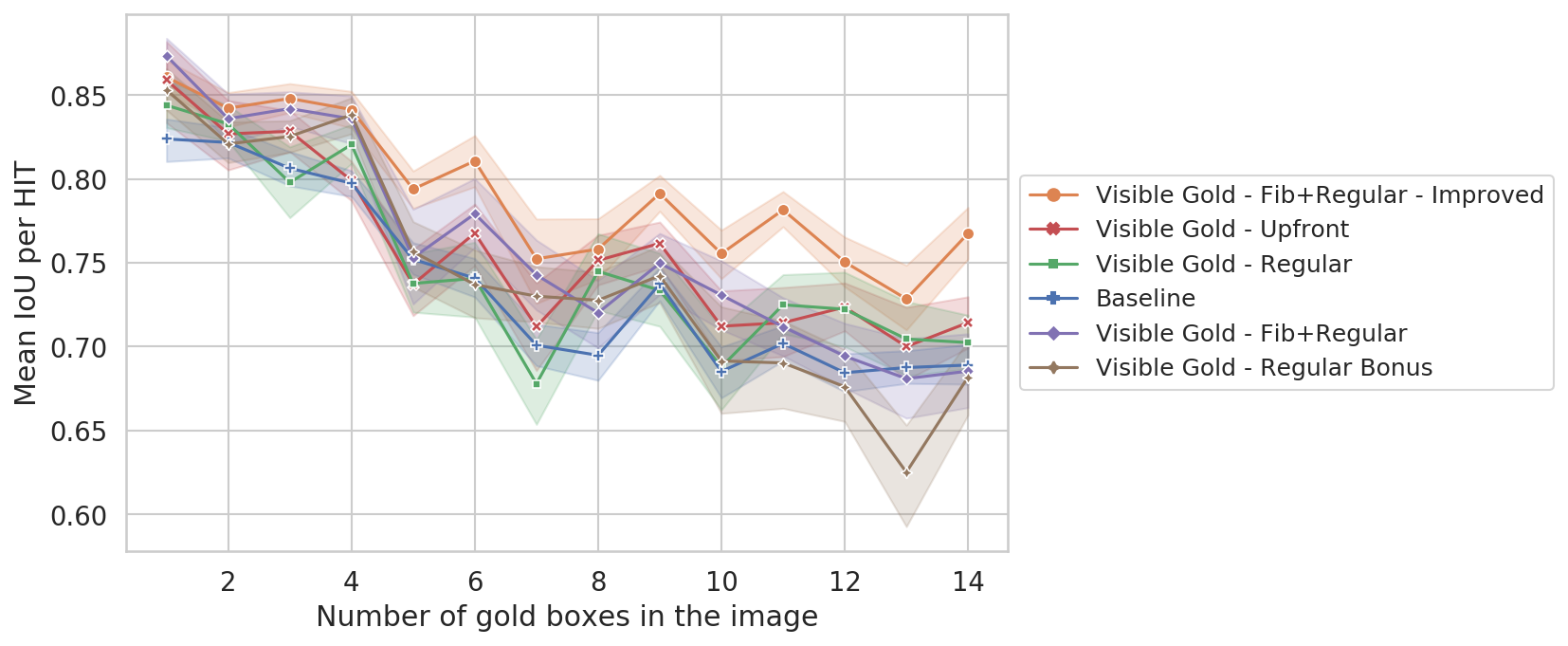}
      \caption{}
      \label{fig:mIoU-workload}
  \end{subfigure}
  \caption{Variation in output quality against task effort for different visible gold conditions. \rev{Shaded areas correspond to standard error.}}
  \label{fig:quality-workload}
\end{figure}

In Figure~\ref{fig:quality-workload}\ml{larger text labels, "ground-truth"->"gold"}, we revisit how the output quality varies when the task effort increases.
We see that our improved visible gold method consistently outperforms other visible gold methods and the baseline in terms of mIoU (Figure~\ref{fig:mIoU-workload}) and recall at mIoU>0.5 (Figure~\ref{fig:misses-workload}).

\begin{figure}[htb]
  \centering
  \includegraphics[width=0.55\linewidth]{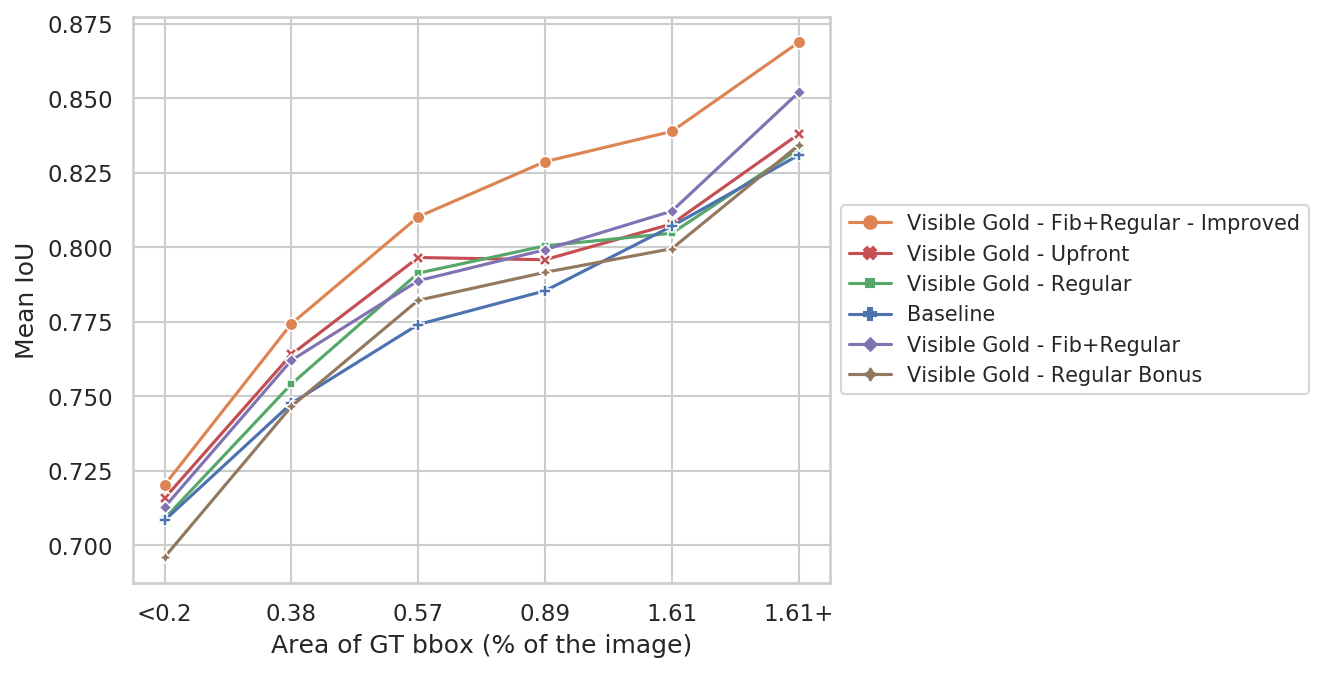}
  \caption{Variation in mIoU against object size for different visible gold conditions.}
  \label{fig:mIoU-target-size}
\end{figure}

Figure~\ref{fig:mIoU-target-size} shows how mIoU varies depending on the size of each ground truth bounding box.
The worker output quality is relatively low for smaller objects.
While this trend is visible across all the methods, the improved visible gold interface outperformed other methods regardless of the target object size.

\subsection{Analysis of Findings}

We first examined the optimum visible gold execution pattern. Our analysis revealed that visible-gold-fib+regular is superior to standard upfront~\cite{gadiraju2015training} or regular~\cite{le2010ensuring} methods. In addition to producing a marginally better overall mIoU score, the fib+regular method helps us preserve the data quality as workers continue to complete tasks. As seen in Figure~\ref{fig:mIoU-hit-order}, work quality declines as workers complete more tasks in both baseline and visible-gold-upfront conditions. However, in visible-gold-regular and visible-gold-fib+regular work quality remains steady as work continues. Fib+regular method is also more advantageous for jobs with a large number of HITs. For example, when a worker completes 100 HITs, regular pattern issues 20\% visible golds, whereas fib+regular issues only 11\%. In Figure~\ref{fig:workers-hitscompleted}, we also observe that a large portion of workers tend to leave the task after completing several HITs in visible-gold conditions. This positive observation confirms that certain workers left the task as they were confronted with quality checks.

Concerning bonus and warning, our results indicate no significant difference. We used these findings to inform the design of improved method where we used the fib+regular as the base visible gold issuing pattern and incorporated both bonus and warning with a tier-based design.





Table~\ref{tab:summary-results-2} presents the impact of different visible gold presentation strategies on the quality of responses for our object detection task.
Note that there is little variation in average time per task among the different visible gold presentation strategies when compared with the other quality-improvement strategies (Table~\ref{tab:summary-results-1}).
On the other hand, the overall quality of annotations increased markedly once we implemented the improved task interface with different ``tiers'' of performance, suggesting that dynamic feedback with clear and transparent communication about penalties and rewards incentivizes higher quality.
Variation in the cadence of visible gold presentation appears to have had less impact than the improvement in the task interface.

Figure~\ref{fig:quality-workload} demonstrates that the improvement from the improved interface can be primarily attributed to better performance on the tasks requiring the most effort as measured by the number of ground truth boxes in the image.
The dynamic tracking and reporting of the worker's running accuracy score on gold data may have made the impact of a small number of poor annotations on a worker's overall performance more clear, incentivizing increased attention to detail on the more difficult tasks.

\section{Discussion}

In this paper, we investigated how data quality improvement mechanisms perform for tasks that involve variable effort. We evaluated common quality enhancement methods and showed that the visible gold method produced annotations of significantly higher quality. We further refined and evaluated the visible gold method, demonstrating the effectiveness of combining upfront and regular testing patterns. Our results also suggest that workers produce better annotations when confronted with consequences via the visible gold feedback interface. However, there was no detectable difference in how bonuses (for high-quality work) and warnings (for low-quality work) affected output quality in the context of visible gold.

\subsection{Data Quality and Variable Effort Tasks}

Our systematic evaluation shows that both object count and object size can impact the annotation quality in object detection task. Our results are consistent with prior work that shows data quality suffers when tasks involve increased effort~\cite{DasSarma2016,kazai2011search}. 
\rev{While there are numerous other crowdsourcing data quality improvement methods (\eg~work strategies~\cite{Han2020CrowdTasks}, task assignment~\cite{Hettiachchi2020CrowdCog}), in this work, we are primarily interested in methods that can potentially support variable effort crowdsourcing.}

Initially, we hypothesized that data quality improvement methods that aim to either standardize the effort (e.g., task decomposition and iterative improvement) or match the pay according to the effort (e.g., variable pay and post-task bonus) should work better when compared to a baseline average pay scheme. However, as detailed in Table~\ref{tab:summary-results-1} and Figure~\ref{fig:quality-effort-1}, none of these methods were successful in surpassing the baseline. While literature~\cite{Bernstein2010Soylent,little2010exploring} highlights that these methods can improve data quality in specific tasks and scenarios, our work shows that such improvements may not hold when the task effort varies. Prior work also highlights that static workflows perform poorly for complex crowd tasks~\cite{retelny2017no}. Therefore, it is important to consider the task effort when using crowdsourcing to obtain annotations as well as when evaluating methods related to crowd work on tasks like bounding box annotation. 

However, we show that out of the detailed methods, visible gold is the most useful method in preserving data quality in variable effort tasks.

\subsection{Visible Gold for Training}

When creating the visible gold mechanism for object detection, we considered three dimensions highlighted in the crowdsourcing literature on worker feedback~\cite{Dow2012}.
In terms of timeliness, we designed our feedback to be synchronous.
However, to prevent workers from guessing which tasks were visible gold questions, they were shown feedback only \textit{after} completing the entire task, i.e., drawing all the bounding boxes for a particular image.
Regarding specificity, we provided automated yet detailed feedback (Figure~\ref{fig:visible-gold-interface}), including aggregate metrics on the image level and fine-grained metrics for each bounding box.

Our study extends prior work on visible gold~\cite{gadiraju2015training,le2010ensuring} by integrating existing testing patterns into a hybrid pattern (Fib+Regular) with both upfront and regular testing.
Our results demonstrate that this hybrid pattern is more effective at maintaining annotation quality over the course of large amounts of tasks.
The capacity to sustain data quality is particularly important as crowd work typically follows a power-law distribution where only a few workers self-select to complete the majority of work, whereas the remainder of the workforce abandons a task early on~\cite{rogstadius2011assessment}. 
Further, our results in Figures~\ref{fig:mIoU-target-size}~ and ~\ref{fig:quality-workload} show that the improved visible gold mechanism is robust when the task involves variable effort with respect to object count and object size.
\rev{Our findings are in line with prior work that uses periodic bonus payments~\cite{Difallah2014Scaling-upImprovement} and achievement priming~\cite{Gadiraju2017ImprovingMicrotasks} to motivate high quality crowd work.}

In addition to improving annotation quality, we argue that our visible gold method possesses a variety of positive attributes from a worker's perspective. First, our method provides transparency around the expected annotation \textit{accuracy}. Second, it provides \rev{task \textit{clarity}~\cite{Gadiraju2017}} to workers by demonstrating how a task should be done by means of concrete examples.
Third, visible gold provides \textit{feedback} to workers, helping them understand their individual task performance, correct specific mistakes, and improve their subsequent work. 
\rev{These three factors contribute to a better understanding of expected task outcomes and reduce the possibility of workers abandoning tasks~\cite{han2019all}, leading to unpaid work.}
Finally, visible gold provides the opportunity for workers to give feedback to the requester if gold standard annotations are faulty or if evaluation results seem to be incorrect (e.g., if there is a bug in the evaluation metric code), which is not possible with hidden gold.

\subsection{Implementing Visible Gold}

Our visible gold method can be easily implemented by task requesters or crowdsourcing platform itself.
We discuss several factors that should be carefully considered in a practical implementation.

\textbf{Generalizability:}
We anticipate that it is possible to develop visible gold templates for many other tasks.
Prior work demonstrates how this can be achieved for text-based classification tasks~\cite{le2010ensuring,gadiraju2015training} and content creation tasks~\cite{gadiraju2015training}.
It would be straightforward to extend the current work for certain tasks like semantic segmentation and keypoint annotation~\cite{Kovashka2016Crowdsourcing} but would require additional effort if the task has no objective evaluation metric like accuracy or mIoU.
While it is trivial to present feedback based on visible golds for object detection, additional explanations would be desired for certain complex tasks.
Future work can explore how to augment visible golds with explanations or rationales.

In the experiment, we picked threshold values for bonuses and blocks based on percentiles in the baseline data.
When implementing visible gold in a crowdsourcing platform, requesters can either set these values based on absolute quality expectations or calibrate thresholds based on initial insights from pilot jobs.

\textbf{Generating Gold:}
As visible golds serve as training examples, it is also important to create a reliable set of visible gold standard questions.
Like hidden gold questions, visible gold should be representative of the dataset and should sufficiently cover edge cases and ambiguous tasks.
While our study assumes that there exists a single objective ground truth answer for any given task, in practice, many tasks are ambiguous~\cite{Schaekermann2018}.
Future work may test evaluation strategies that accept multiple valid ground truth answers~\cite{chang2017revolt,Schaekermann2020}.
The problem of generating high-quality gold standards is a non-trivial process and remains an open research challenge.
An important challenge for future work is how visible gold mechanisms can work when available ground truth data are imperfect or noisy.

\textbf{Presenting Feedback:}
The visible gold presentation in the current work includes detailed feedback for each visible gold answer, providing an overall accuracy score along with metrics and visual feedback for each work unit.
We also added continuous feedback on quality checks during further refinements.
However, we identify several future improvements for visible gold interfaces.
First, future designs may introduce a `revise and resubmit' mechanism.
The design implemented in this study did not let workers adjust their original annotation after being presented with feedback on a visible gold question.
Correcting their original annotation according to the gold answer could help workers understand how to achieve higher task accuracy through experiential learning.
Second, the interface could be enhanced by adding interactive feedback. In the current implementation, workers only receive feedback on their work once they submit the answer. It is also possible to explore whether interactive feedback for partially completed tasks is helpful for the workers. This can be more meaningful for highly complex tasks such as bounding box or 3D point annotation tasks with a large number of target objects in the same image/task. Third, similar to worker-led instruction refinement~\cite{Manam2019TaskMate} or workflow creation~\cite{Kulkarni2012Collaboratively}, we could encourage workers to provide feedback on visible gold. If a worker encounters a flawed visible gold question, there should be a way to flag it or provide detailed feedback. This way, requesters can remove the reported visible gold questions from the task. As worker quality is measured through visible golds, this is an important enhancement when implementing visible gold at scale. In addition to the information provided regarding work quality, we could explore interventions such as micro-diversions~\cite{dai2015and}, dedicated training sub-tasks~\cite{doroudi2016toward}, mandatory instruction documents, etc.

\textbf{Testing Patterns:}
The adaptive gold execution strategy introduced in this work can be expanded by considering worker quality metrics outside the current job.
For instance, if we already know a particular worker is doing well in object detection based on previous jobs, we can reduce the visible gold frequency.
It is also possible to draw from prior work that investigates how to issue hidden gold questions optimally~\cite{Bragg2016OptimalWorkers}.
Finally, the positive impact of the tiering system in the improved task interface suggests two interesting directions for future study: (1) whether we can expand it to a platform-level system and (2) how we can estimate/dynamically vary threshold quality values and rewards to improve performance. 

\tm{Does the dynamic nature of the task serving in your setup help with this problem for our experiment at least Danula? Added commented text below}The use of visible golds also has several inherent limitations that practitioners should be aware of. With visible golds, workers can easily flag the HITs as Gold Questions, and with the help of third-party plugins, other workers may be able to detect in advance whether a particular HIT is gold or not~\cite{checco2020adversarial,checco2018all}.


\subsection{Limitations}

We acknowledge several limitations of our study. First, in our crowdsourcing experiments, workers were allowed to attempt an arbitrary number of questions instead of being assigned a fixed quota. We made this design decision to match the typical workflow in crowdsourcing platforms and therefore ensure the ecological validity of our work. \tm{I added a comment on the effect of allowing an arbitrary number of responses --- can we say that it may have reduced the statistical power by increasing variance of our statistics?} As a result, the distribution of work between workers was uneven. Second, we did not specifically collect worker demographic information that may impact the output quality. However, we utilize a curated worker pool that excludes workers who would intentionally spam the task or produce low-quality data.

\section{Conclusion}

In this paper, we systematically evaluated the impact of existing quality improvement methods for tasks involving variable effort.
Our results from a series of crowdsourced experiments in the context of object detection show that providing feedback to human annotators via visible gold can produce better quality outcomes than methods aiming to balance effort and pay at the item level through adjusting pay per item or decomposing the task into chunks of similar effort.
We further designed and empirically evaluated variants of the visible gold method testing different issuance patterns and quality-related consequences.
Our final design iteration of visible gold combined dynamic testing patterns with tier-based consequences and significantly improved bounding box accuracy by 5.7\% compared to a basic visible gold variant and by 7.5\% compared to a baseline without visible gold.
Our work broadens the understanding of quality assurance processes for variable effort annotation tasks and emphasizes the value of visible gold-based feedback mechanisms in this process.

\begin{acks}
We thank the many talented Amazon Mechanical Turk workers who contributed to our study and made this work possible. We also thank our reviewers for their valuable feedback. We further acknowledge other members of the human-in-the-loop (HIL) science team for their valuable comments. Any opinions, findings, and conclusions or recommendations expressed by the authors are entirely their own.
\end{acks}

\bibliographystyle{ACM-Reference-Format}
\bibliography{ref}


\begin{thebibliography}{86}


\ifx \showCODEN    \undefined \def \showCODEN     #1{\unskip}     \fi
\ifx \showDOI      \undefined \def \showDOI       #1{#1}\fi
\ifx \showISBNx    \undefined \def \showISBNx     #1{\unskip}     \fi
\ifx \showISBNxiii \undefined \def \showISBNxiii  #1{\unskip}     \fi
\ifx \showISSN     \undefined \def \showISSN      #1{\unskip}     \fi
\ifx \showLCCN     \undefined \def \showLCCN      #1{\unskip}     \fi
\ifx \shownote     \undefined \def \shownote      #1{#1}          \fi
\ifx \showarticletitle \undefined \def \showarticletitle #1{#1}   \fi
\ifx \showURL      \undefined \def \showURL       {\relax}        \fi
\providecommand\bibfield[2]{#2}
\providecommand\bibinfo[2]{#2}
\providecommand\natexlab[1]{#1}
\providecommand\showeprint[2][]{arXiv:#2}

\bibitem[\protect\citeauthoryear{Adhikari and Huttunen}{Adhikari and
  Huttunen}{2020}]%
        {adhikari2020iterative}
\bibfield{author}{\bibinfo{person}{Bishwo Adhikari} {and}
  \bibinfo{person}{Heikki Huttunen}.} \bibinfo{year}{2020}\natexlab{}.
\newblock \showarticletitle{Iterative Bounding Box Annotation for Object
  Detection}. In \bibinfo{booktitle}{\emph{International Conference on Pattern
  Recognition (ICPR)}}.
\newblock


\bibitem[\protect\citeauthoryear{Adhikari, Peltomaki, Puura, and
  Huttunen}{Adhikari et~al\mbox{.}}{2018}]%
        {adhikari2018faster}
\bibfield{author}{\bibinfo{person}{Bishwo Adhikari}, \bibinfo{person}{Jukka
  Peltomaki}, \bibinfo{person}{Jussi Puura}, {and} \bibinfo{person}{Heikki
  Huttunen}.} \bibinfo{year}{2018}\natexlab{}.
\newblock \showarticletitle{Faster Bounding Box Annotation for Object Detection
  in Indoor Scenes}. In \bibinfo{booktitle}{\emph{2018 7th European Workshop on
  Visual Information Processing (EUVIP)}}. \bibinfo{pages}{1--6}.
\newblock
\urldef\tempurl%
\url{https://doi.org/10.1109/EUVIP.2018.8611732}
\showDOI{\tempurl}


\bibitem[\protect\citeauthoryear{Aipe and Gadiraju}{Aipe and Gadiraju}{2018}]%
        {Aipe2018}
\bibfield{author}{\bibinfo{person}{Alan Aipe} {and} \bibinfo{person}{Ujwal
  Gadiraju}.} \bibinfo{year}{2018}\natexlab{}.
\newblock \showarticletitle{{SimilarHITs: Revealing the Role of Task Similarity
  in Microtask Crowdsourcing}}. In \bibinfo{booktitle}{\emph{Proceedings of the
  29th on Hypertext and Social Media}} \emph{(\bibinfo{series}{HT '18})}.
  \bibinfo{publisher}{ACM}, \bibinfo{address}{New York, NY, USA},
  \bibinfo{pages}{115–122}.
\newblock
\showISBNx{9781450354271}
\urldef\tempurl%
\url{https://doi.org/10.1145/3209542.3209558}
\showDOI{\tempurl}


\bibitem[\protect\citeauthoryear{Ambati, Vogel, and Carbonell}{Ambati
  et~al\mbox{.}}{2012}]%
        {Ambati2012Collaborative}
\bibfield{author}{\bibinfo{person}{Vamshi Ambati}, \bibinfo{person}{Stephan
  Vogel}, {and} \bibinfo{person}{Jaime Carbonell}.}
  \bibinfo{year}{2012}\natexlab{}.
\newblock \showarticletitle{Collaborative Workflow for Crowdsourcing
  Translation}. In \bibinfo{booktitle}{\emph{Proceedings of the ACM 2012
  Conference on Computer Supported Cooperative Work}} (Seattle, Washington,
  USA) \emph{(\bibinfo{series}{CSCW '12})}. \bibinfo{publisher}{ACM},
  \bibinfo{address}{New York, NY, USA}, \bibinfo{pages}{1191–1194}.
\newblock
\showISBNx{9781450310864}
\urldef\tempurl%
\url{https://doi.org/10.1145/2145204.2145382}
\showDOI{\tempurl}


\bibitem[\protect\citeauthoryear{Attenberg, Ipeirotis, and Provost}{Attenberg
  et~al\mbox{.}}{2011}]%
        {attenberg2011beat}
\bibfield{author}{\bibinfo{person}{Josh Attenberg},
  \bibinfo{person}{Panagiotis~G. Ipeirotis}, {and} \bibinfo{person}{Foster
  Provost}.} \bibinfo{year}{2011}\natexlab{}.
\newblock \showarticletitle{Beat the Machine: Challenging Workers to Find the
  Unknown Unknowns}. In \bibinfo{booktitle}{\emph{Proceedings of the AAAI
  Conference on Human Computation}} \emph{(\bibinfo{series}{HCOMP})}.
  \bibinfo{publisher}{AAAI Press}, \bibinfo{pages}{2–7}.
\newblock


\bibitem[\protect\citeauthoryear{Bernstein, Little, Miller, Hartmann, Ackerman,
  Karger, Crowell, and Panovich}{Bernstein et~al\mbox{.}}{2010}]%
        {Bernstein2010Soylent}
\bibfield{author}{\bibinfo{person}{Michael~S. Bernstein}, \bibinfo{person}{Greg
  Little}, \bibinfo{person}{Robert~C. Miller}, \bibinfo{person}{Björn
  Hartmann}, \bibinfo{person}{Mark~S. Ackerman}, \bibinfo{person}{David~R.
  Karger}, \bibinfo{person}{David Crowell}, {and} \bibinfo{person}{Katrina
  Panovich}.} \bibinfo{year}{2010}\natexlab{}.
\newblock \showarticletitle{{Soylent: A Word Processor with a Crowd Inside}}.
  In \bibinfo{booktitle}{\emph{Proceedings of the 23nd Annual ACM Symposium on
  User Interface Software and Technology}} \emph{(\bibinfo{series}{UIST '10})}.
  \bibinfo{publisher}{ACM}, \bibinfo{address}{New York, NY, USA},
  \bibinfo{pages}{313--322}.
\newblock
\showISBNx{978-1-4503-0271-5}
\urldef\tempurl%
\url{https://doi.org/10.1145/1866029.1866078}
\showDOI{\tempurl}


\bibitem[\protect\citeauthoryear{Borji, Cheng, Hou, Jiang, and Li}{Borji
  et~al\mbox{.}}{2019}]%
        {borji2019salient}
\bibfield{author}{\bibinfo{person}{Ali Borji}, \bibinfo{person}{Ming-Ming
  Cheng}, \bibinfo{person}{Qibin Hou}, \bibinfo{person}{Huaizu Jiang}, {and}
  \bibinfo{person}{Jia Li}.} \bibinfo{year}{2019}\natexlab{}.
\newblock \showarticletitle{Salient object detection: A survey}.
\newblock \bibinfo{journal}{\emph{Computational Visual Media}}
  \bibinfo{volume}{5}, \bibinfo{number}{2} (\bibinfo{year}{2019}),
  \bibinfo{pages}{117--150}.
\newblock
\showISBNx{2096-0662}
\urldef\tempurl%
\url{https://doi.org/10.1007/s41095-019-0149-9}
\showDOI{\tempurl}


\bibitem[\protect\citeauthoryear{Bragg, {Mausam}, and Weld}{Bragg
  et~al\mbox{.}}{2016}]%
        {Bragg2016OptimalWorkers}
\bibfield{author}{\bibinfo{person}{Jonathan Bragg}, \bibinfo{person}{{Mausam}},
  {and} \bibinfo{person}{Daniel~S. Weld}.} \bibinfo{year}{2016}\natexlab{}.
\newblock \showarticletitle{{Optimal testing for crowd workers}}.
\newblock \bibinfo{journal}{\emph{Proceedings of the International Joint
  Conference on Autonomous Agents and Multiagent Systems, AAMAS}}
  (\bibinfo{year}{2016}), \bibinfo{pages}{966--974}.
\newblock
\showISBNx{9781450342391}
\showISSN{15582914}


\bibitem[\protect\citeauthoryear{Buhrmester, Kwang, and Gosling}{Buhrmester
  et~al\mbox{.}}{2011}]%
        {buhrmester2011amazon}
\bibfield{author}{\bibinfo{person}{Michael Buhrmester}, \bibinfo{person}{Tracy
  Kwang}, {and} \bibinfo{person}{Samuel~D Gosling}.}
  \bibinfo{year}{2011}\natexlab{}.
\newblock \showarticletitle{Amazon’s Mechanical Turk: A New Source of
  Inexpensive, Yet High-Quality, Data?}
\newblock \bibinfo{journal}{\emph{Perspectives on Psychological Science}}
  \bibinfo{volume}{6}, \bibinfo{number}{1} (\bibinfo{year}{2011}),
  \bibinfo{pages}{3--5}.
\newblock


\bibitem[\protect\citeauthoryear{Cai, Iqbal, and Teevan}{Cai
  et~al\mbox{.}}{2016}]%
        {Cai2016}
\bibfield{author}{\bibinfo{person}{Carrie~J Cai}, \bibinfo{person}{Shamsi~T
  Iqbal}, {and} \bibinfo{person}{Jaime Teevan}.}
  \bibinfo{year}{2016}\natexlab{}.
\newblock \showarticletitle{{Chain Reactions: The Impact of Order on Microtask
  Chains}}. In \bibinfo{booktitle}{\emph{Proceedings of the 2016 CHI Conference
  on Human Factors in Computing Systems}} \emph{(\bibinfo{series}{CHI '16})}.
  \bibinfo{publisher}{ACM}, \bibinfo{address}{New York, NY, USA},
  \bibinfo{pages}{3143–3154}.
\newblock
\showISBNx{9781450333627}
\urldef\tempurl%
\url{https://doi.org/10.1145/2858036.2858237}
\showDOI{\tempurl}


\bibitem[\protect\citeauthoryear{Chang, Amershi, and Kamar}{Chang
  et~al\mbox{.}}{2017}]%
        {chang2017revolt}
\bibfield{author}{\bibinfo{person}{Joseph~Chee Chang}, \bibinfo{person}{Saleema
  Amershi}, {and} \bibinfo{person}{Ece Kamar}.}
  \bibinfo{year}{2017}\natexlab{}.
\newblock \showarticletitle{Revolt: Collaborative Crowdsourcing for Labeling
  Machine Learning Datasets}. In \bibinfo{booktitle}{\emph{Proceedings of the
  2017 CHI Conference on Human Factors in Computing Systems}} (Denver,
  Colorado, USA) \emph{(\bibinfo{series}{CHI '17})}. \bibinfo{publisher}{ACM},
  \bibinfo{address}{New York, NY, USA}, \bibinfo{pages}{2334–2346}.
\newblock
\showISBNx{9781450346559}
\urldef\tempurl%
\url{https://doi.org/10.1145/3025453.3026044}
\showDOI{\tempurl}


\bibitem[\protect\citeauthoryear{Checco, Bates, and Demartini}{Checco
  et~al\mbox{.}}{2018}]%
        {checco2018all}
\bibfield{author}{\bibinfo{person}{Alessandro Checco}, \bibinfo{person}{Jo
  Bates}, {and} \bibinfo{person}{Gianluca Demartini}.}
  \bibinfo{year}{2018}\natexlab{}.
\newblock \showarticletitle{All That Glitters is Gold -- An Attack Scheme on
  Gold Questions in Crowdsourcing}. In \bibinfo{booktitle}{\emph{Proceedings of
  the Sixth AAAI Conference on Human Computation and Crowdsourcing}}
  \emph{(\bibinfo{series}{HCOMP})}. AAAI Press.
\newblock


\bibitem[\protect\citeauthoryear{Checco, Bates, and Demartini}{Checco
  et~al\mbox{.}}{2020}]%
        {checco2020adversarial}
\bibfield{author}{\bibinfo{person}{Alessandro Checco}, \bibinfo{person}{Jo
  Bates}, {and} \bibinfo{person}{Gianluca Demartini}.}
  \bibinfo{year}{2020}\natexlab{}.
\newblock \showarticletitle{Adversarial Attacks on Crowdsourcing Quality
  Control}.
\newblock \bibinfo{journal}{\emph{Journal of Artificial Intelligence Research}}
   \bibinfo{volume}{67} (\bibinfo{year}{2020}), \bibinfo{pages}{375--408}.
\newblock
\urldef\tempurl%
\url{https://doi.org/10.1613/jair.1.11332}
\showDOI{\tempurl}


\bibitem[\protect\citeauthoryear{Cheng, Teevan, and Bernstein}{Cheng
  et~al\mbox{.}}{2015}]%
        {cheng2015measuring}
\bibfield{author}{\bibinfo{person}{Justin Cheng}, \bibinfo{person}{Jaime
  Teevan}, {and} \bibinfo{person}{Michael~S. Bernstein}.}
  \bibinfo{year}{2015}\natexlab{}.
\newblock \showarticletitle{Measuring Crowdsourcing Effort with Error-Time
  Curves}. In \bibinfo{booktitle}{\emph{Proceedings of the 33rd Annual ACM
  Conference on Human Factors in Computing Systems}} (Seoul, Republic of Korea)
  \emph{(\bibinfo{series}{CHI '15})}. \bibinfo{publisher}{ACM},
  \bibinfo{address}{New York, NY, USA}, \bibinfo{pages}{1365–1374}.
\newblock
\showISBNx{9781450331456}
\urldef\tempurl%
\url{https://doi.org/10.1145/2702123.2702145}
\showDOI{\tempurl}


\bibitem[\protect\citeauthoryear{Dai, Lin, Mausam, and Weld}{Dai
  et~al\mbox{.}}{2013}]%
        {dai2013pomdp}
\bibfield{author}{\bibinfo{person}{Peng Dai}, \bibinfo{person}{Christopher~H.
  Lin}, \bibinfo{person}{Mausam}, {and} \bibinfo{person}{Daniel~S. Weld}.}
  \bibinfo{year}{2013}\natexlab{}.
\newblock \showarticletitle{POMDP-based control of workflows for
  crowdsourcing}.
\newblock \bibinfo{journal}{\emph{Artificial Intelligence}}
  \bibinfo{volume}{202} (\bibinfo{year}{2013}), \bibinfo{pages}{52 -- 85}.
\newblock
\showISSN{0004-3702}
\urldef\tempurl%
\url{https://doi.org/10.1016/j.artint.2013.06.002}
\showDOI{\tempurl}


\bibitem[\protect\citeauthoryear{Dai, Rzeszotarski, Paritosh, and Chi}{Dai
  et~al\mbox{.}}{2015}]%
        {dai2015and}
\bibfield{author}{\bibinfo{person}{Peng Dai}, \bibinfo{person}{Jeffrey~M.
  Rzeszotarski}, \bibinfo{person}{Praveen Paritosh}, {and}
  \bibinfo{person}{Ed~H. Chi}.} \bibinfo{year}{2015}\natexlab{}.
\newblock \showarticletitle{And Now for Something Completely Different:
  Improving Crowdsourcing Workflows with Micro-Diversions}. In
  \bibinfo{booktitle}{\emph{Proceedings of the 18th ACM Conference on Computer
  Supported Cooperative Work \& Social Computing}} (Vancouver, BC, Canada)
  \emph{(\bibinfo{series}{CSCW '15})}. \bibinfo{publisher}{ACM},
  \bibinfo{address}{New York, NY, USA}, \bibinfo{pages}{628–638}.
\newblock
\showISBNx{9781450329224}
\urldef\tempurl%
\url{https://doi.org/10.1145/2675133.2675260}
\showDOI{\tempurl}


\bibitem[\protect\citeauthoryear{Daniel, Kucherbaev, Cappiello, Benatallah, and
  Allahbakhsh}{Daniel et~al\mbox{.}}{2018}]%
        {Daniel2018}
\bibfield{author}{\bibinfo{person}{Florian Daniel}, \bibinfo{person}{Pavel
  Kucherbaev}, \bibinfo{person}{Cinzia Cappiello}, \bibinfo{person}{Boualem
  Benatallah}, {and} \bibinfo{person}{Mohammad Allahbakhsh}.}
  \bibinfo{year}{2018}\natexlab{}.
\newblock \showarticletitle{{Quality Control in Crowdsourcing: A Survey of
  Quality Attributes, Assessment Techniques, and Assurance Actions}}.
\newblock \bibinfo{journal}{\emph{Comput. Surveys}} \bibinfo{volume}{51},
  \bibinfo{number}{1} (\bibinfo{date}{4} \bibinfo{year}{2018}),
  \bibinfo{pages}{1--40}.
\newblock
\showISSN{0360-0300}
\urldef\tempurl%
\url{https://doi.org/10.1145/3148148}
\showDOI{\tempurl}


\bibitem[\protect\citeauthoryear{Das~Sarma, Parameswaran, and Widom}{Das~Sarma
  et~al\mbox{.}}{2016}]%
        {DasSarma2016}
\bibfield{author}{\bibinfo{person}{Akash Das~Sarma}, \bibinfo{person}{Aditya
  Parameswaran}, {and} \bibinfo{person}{Jennifer Widom}.}
  \bibinfo{year}{2016}\natexlab{}.
\newblock \showarticletitle{{Towards Globally Optimal Crowdsourcing Quality
  Management}}. In \bibinfo{booktitle}{\emph{Proceedings of the 2016
  International Conference on Management of Data - SIGMOD '16}}
  \emph{(\bibinfo{series}{SIGMOD '16}, Vol.~\bibinfo{volume}{26-June-20})}.
  \bibinfo{publisher}{ACM Press}, \bibinfo{address}{New York, New York, USA},
  \bibinfo{pages}{47--62}.
\newblock
\showISBNx{9781450335317}
\showISSN{07308078}
\urldef\tempurl%
\url{https://doi.org/10.1145/2882903.2882953}
\showDOI{\tempurl}


\bibitem[\protect\citeauthoryear{Difallah, Catasta, Demartini, Cudr, and
  Cudr{\'{e}}-Mauroux}{Difallah et~al\mbox{.}}{2014}]%
        {Difallah2014Scaling-upImprovement}
\bibfield{author}{\bibinfo{person}{Djellel~Eddine Difallah},
  \bibinfo{person}{Michele Catasta}, \bibinfo{person}{Gianluca Demartini},
  \bibinfo{person}{Philippe Cudr}, {and} \bibinfo{person}{Philippe
  Cudr{\'{e}}-Mauroux}.} \bibinfo{year}{2014}\natexlab{}.
\newblock \showarticletitle{{Scaling-up the Crowd: Micro-Task Pricing Schemes
  for Worker Retention and Latency Improvement}}.
\newblock \bibinfo{journal}{\emph{Second AAAI Conference on Human Computation
  and Crowdsourcing}} \bibinfo{number}{Hcomp} (\bibinfo{year}{2014}),
  \bibinfo{pages}{50--58}.
\newblock


\bibitem[\protect\citeauthoryear{Doroudi, Kamar, Brunskill, and
  Horvitz}{Doroudi et~al\mbox{.}}{2016}]%
        {doroudi2016toward}
\bibfield{author}{\bibinfo{person}{Shayan Doroudi}, \bibinfo{person}{Ece
  Kamar}, \bibinfo{person}{Emma Brunskill}, {and} \bibinfo{person}{Eric
  Horvitz}.} \bibinfo{year}{2016}\natexlab{}.
\newblock \showarticletitle{Toward a Learning Science for Complex Crowdsourcing
  Tasks}. In \bibinfo{booktitle}{\emph{Proceedings of the 2016 CHI Conference
  on Human Factors in Computing Systems}} (San Jose, California, USA)
  \emph{(\bibinfo{series}{CHI '16})}. \bibinfo{publisher}{ACM},
  \bibinfo{address}{New York, NY, USA}, \bibinfo{pages}{2623–2634}.
\newblock
\showISBNx{9781450333627}
\urldef\tempurl%
\url{https://doi.org/10.1145/2858036.2858268}
\showDOI{\tempurl}


\bibitem[\protect\citeauthoryear{Dow, Kulkarni, Klemmer, and Hartmann}{Dow
  et~al\mbox{.}}{2012}]%
        {Dow2012}
\bibfield{author}{\bibinfo{person}{Steven~P. Dow}, \bibinfo{person}{Anand
  Kulkarni}, \bibinfo{person}{Scott Klemmer}, {and} \bibinfo{person}{Björn
  Hartmann}.} \bibinfo{year}{2012}\natexlab{}.
\newblock \showarticletitle{{Shepherding the Crowd Yields Better Work}}. In
  \bibinfo{booktitle}{\emph{Proceedings of the ACM 2012 Conference on Computer
  Supported Cooperative Work}} \emph{(\bibinfo{series}{CSCW '12})}.
  \bibinfo{publisher}{ACM}, \bibinfo{address}{New York, NY, USA},
  \bibinfo{pages}{1013--1022}.
\newblock
\showISBNx{978-1-4503-1086-4}
\urldef\tempurl%
\url{https://doi.org/10.1145/2145204.2145355}
\showDOI{\tempurl}


\bibitem[\protect\citeauthoryear{Fan, Li, Ooi, Tan, and Feng}{Fan
  et~al\mbox{.}}{2015}]%
        {Fan2015}
\bibfield{author}{\bibinfo{person}{Ju Fan}, \bibinfo{person}{Guoliang Li},
  \bibinfo{person}{Beng~Chin Ooi}, \bibinfo{person}{Kian-lee Tan}, {and}
  \bibinfo{person}{Jianhua Feng}.} \bibinfo{year}{2015}\natexlab{}.
\newblock \showarticletitle{{iCrowd: An Adaptive Crowdsourcing Framework}}. In
  \bibinfo{booktitle}{\emph{Proceedings of the 2015 ACM SIGMOD International
  Conference on Management of Data}} \emph{(\bibinfo{series}{SIGMOD '15})}.
  \bibinfo{publisher}{ACM}, \bibinfo{address}{New York, NY, USA},
  \bibinfo{pages}{1015--1030}.
\newblock
\showISBNx{978-1-4503-2758-9}
\urldef\tempurl%
\url{https://doi.org/10.1145/2723372.2750550}
\showDOI{\tempurl}


\bibitem[\protect\citeauthoryear{Gadiraju and Dietze}{Gadiraju and
  Dietze}{2017}]%
        {Gadiraju2017ImprovingMicrotasks}
\bibfield{author}{\bibinfo{person}{Ujwal Gadiraju} {and}
  \bibinfo{person}{Stefan Dietze}.} \bibinfo{year}{2017}\natexlab{}.
\newblock \showarticletitle{{Improving Learning through Achievement Priming in
  Crowdsourced Information Finding Microtasks}}. In
  \bibinfo{booktitle}{\emph{Proceedings of the Seventh International Learning
  Analytics {\&}amp; Knowledge Conference}} \emph{(\bibinfo{series}{LAK '17})}.
  \bibinfo{publisher}{ACM}, \bibinfo{address}{New York, NY, USA},
  \bibinfo{pages}{105–114}.
\newblock
\showISBNx{9781450348706}
\urldef\tempurl%
\url{https://doi.org/10.1145/3027385.3027402}
\showDOI{\tempurl}


\bibitem[\protect\citeauthoryear{Gadiraju, Fetahu, and Kawase}{Gadiraju
  et~al\mbox{.}}{2015}]%
        {gadiraju2015training}
\bibfield{author}{\bibinfo{person}{Ujwal Gadiraju}, \bibinfo{person}{Besnik
  Fetahu}, {and} \bibinfo{person}{Ricardo Kawase}.}
  \bibinfo{year}{2015}\natexlab{}.
\newblock \showarticletitle{Training Workers for Improving Performance in
  Crowdsourcing Microtasks}.
\newblock In \bibinfo{booktitle}{\emph{Design for Teaching and Learning in a
  Networked World}}. \bibinfo{publisher}{Springer}, \bibinfo{address}{Cham},
  \bibinfo{pages}{100--114}.
\newblock
\urldef\tempurl%
\url{https://doi.org/10.1007/978-3-319-24258-3_8}
\showDOI{\tempurl}


\bibitem[\protect\citeauthoryear{Gadiraju, Yang, and Bozzon}{Gadiraju
  et~al\mbox{.}}{2017}]%
        {Gadiraju2017}
\bibfield{author}{\bibinfo{person}{Ujwal Gadiraju}, \bibinfo{person}{Jie Yang},
  {and} \bibinfo{person}{Alessandro Bozzon}.} \bibinfo{year}{2017}\natexlab{}.
\newblock \showarticletitle{{Clarity is a Worthwhile Quality: On the Role of
  Task Clarity in Microtask Crowdsourcing}}. In
  \bibinfo{booktitle}{\emph{Proceedings of the 28th ACM Conference on Hypertext
  and Social Media - HT '17}} \emph{(\bibinfo{series}{HT '17})}.
  \bibinfo{publisher}{ACM Press}, \bibinfo{address}{New York, New York, USA},
  \bibinfo{pages}{5--14}.
\newblock
\showISBNx{9781450347082}
\urldef\tempurl%
\url{https://doi.org/10.1145/3078714.3078715}
\showDOI{\tempurl}


\bibitem[\protect\citeauthoryear{Goto, Ishida, and Lin}{Goto
  et~al\mbox{.}}{2016}]%
        {goto2016understanding}
\bibfield{author}{\bibinfo{person}{Shinsuke Goto}, \bibinfo{person}{Toru
  Ishida}, {and} \bibinfo{person}{Donghui Lin}.}
  \bibinfo{year}{2016}\natexlab{}.
\newblock \showarticletitle{Understanding crowdsourcing workflow: modeling and
  optimizing iterative and parallel processes}. In
  \bibinfo{booktitle}{\emph{Proceedings of the Fourth AAAI Conference on Human
  Computation and Crowdsourcing}} \emph{(\bibinfo{series}{HCOMP},
  Vol.~\bibinfo{volume}{4})}. \bibinfo{publisher}{AAAI Press}.
\newblock


\bibitem[\protect\citeauthoryear{Grady and Lease}{Grady and Lease}{2010}]%
        {Grady10}
\bibfield{author}{\bibinfo{person}{Catherine Grady} {and}
  \bibinfo{person}{Matthew Lease}.} \bibinfo{year}{2010}\natexlab{}.
\newblock \showarticletitle{Crowdsourcing Document Relevance Assessment with
  Mechanical Turk}. In \bibinfo{booktitle}{\emph{Proceedings of the NAACL HLT
  2010 Workshop on Creating Speech and Language Data with Amazon's Mechanical
  Turk}}. \bibinfo{publisher}{Association for Computational Linguistics},
  \bibinfo{address}{Los Angeles}, \bibinfo{pages}{172--179}.
\newblock


\bibitem[\protect\citeauthoryear{Han, Maddalena, Checco, Sarasua, Gadiraju,
  Roitero, and Demartini}{Han et~al\mbox{.}}{2020}]%
        {Han2020CrowdTasks}
\bibfield{author}{\bibinfo{person}{Lei Han}, \bibinfo{person}{Eddy Maddalena},
  \bibinfo{person}{Alessandro Checco}, \bibinfo{person}{Cristina Sarasua},
  \bibinfo{person}{Ujwal Gadiraju}, \bibinfo{person}{Kevin Roitero}, {and}
  \bibinfo{person}{Gianluca Demartini}.} \bibinfo{year}{2020}\natexlab{}.
\newblock \showarticletitle{{Crowd worker strategies in relevance judgment
  tasks}}. In \bibinfo{booktitle}{\emph{WSDM 2020 - Proceedings of the 13th
  International Conference on Web Search and Data Mining}}.
\newblock
\showISBNx{9781450368223}
\urldef\tempurl%
\url{https://doi.org/10.1145/3336191.3371857}
\showDOI{\tempurl}


\bibitem[\protect\citeauthoryear{Han, Roitero, Gadiraju, Sarasua, Checco,
  Maddalena, and Demartini}{Han et~al\mbox{.}}{2019}]%
        {han2019all}
\bibfield{author}{\bibinfo{person}{Lei Han}, \bibinfo{person}{Kevin Roitero},
  \bibinfo{person}{Ujwal Gadiraju}, \bibinfo{person}{Cristina Sarasua},
  \bibinfo{person}{Alessandro Checco}, \bibinfo{person}{Eddy Maddalena}, {and}
  \bibinfo{person}{Gianluca Demartini}.} \bibinfo{year}{2019}\natexlab{}.
\newblock \showarticletitle{All Those Wasted Hours: On Task Abandonment in
  Crowdsourcing}. In \bibinfo{booktitle}{\emph{Proceedings of the Twelfth ACM
  International Conference on Web Search and Data Mining}} (Melbourne VIC,
  Australia) \emph{(\bibinfo{series}{WSDM '19})}. \bibinfo{publisher}{ACM},
  \bibinfo{address}{New York, NY, USA}, \bibinfo{pages}{321–329}.
\newblock
\showISBNx{9781450359405}
\urldef\tempurl%
\url{https://doi.org/10.1145/3289600.3291035}
\showDOI{\tempurl}


\bibitem[\protect\citeauthoryear{Hara, Le, and Froehlich}{Hara
  et~al\mbox{.}}{2013}]%
        {hara2013combining}
\bibfield{author}{\bibinfo{person}{Kotaro Hara}, \bibinfo{person}{Vicki Le},
  {and} \bibinfo{person}{Jon Froehlich}.} \bibinfo{year}{2013}\natexlab{}.
\newblock \showarticletitle{Combining Crowdsourcing and Google Street View to
  Identify Street-Level Accessibility Problems}. In
  \bibinfo{booktitle}{\emph{Proceedings of the SIGCHI Conference on Human
  Factors in Computing Systems}} (Paris, France) \emph{(\bibinfo{series}{CHI
  '13})}. \bibinfo{publisher}{ACM}, \bibinfo{address}{New York, NY, USA},
  \bibinfo{pages}{631–640}.
\newblock
\showISBNx{9781450318990}
\urldef\tempurl%
\url{https://doi.org/10.1145/2470654.2470744}
\showDOI{\tempurl}


\bibitem[\protect\citeauthoryear{Hettiachchi, van Berkel, Kostakos, and
  Goncalves}{Hettiachchi et~al\mbox{.}}{2020}]%
        {Hettiachchi2020CrowdCog}
\bibfield{author}{\bibinfo{person}{Danula Hettiachchi}, \bibinfo{person}{Niels
  van Berkel}, \bibinfo{person}{Vassilis Kostakos}, {and}
  \bibinfo{person}{Jorge Goncalves}.} \bibinfo{year}{2020}\natexlab{}.
\newblock \showarticletitle{{CrowdCog: A Cognitive Skill based System for
  Heterogeneous Task Assignment and Recommendation in Crowdsourcing}}.
\newblock \bibinfo{journal}{\emph{Proceedings of the ACM on Human-Computer
  Interaction}} \bibinfo{volume}{4}, \bibinfo{number}{CSCW2}
  (\bibinfo{date}{10} \bibinfo{year}{2020}), \bibinfo{pages}{1--22}.
\newblock
\showISSN{2573-0142}
\urldef\tempurl%
\url{https://doi.org/10.1145/3415181}
\showDOI{\tempurl}


\bibitem[\protect\citeauthoryear{Ho, Slivkins, Suri, and Vaughan}{Ho
  et~al\mbox{.}}{2015}]%
        {ho2015incentivizing}
\bibfield{author}{\bibinfo{person}{Chien~Ju Ho}, \bibinfo{person}{Aleksandrs
  Slivkins}, \bibinfo{person}{Siddharth Suri}, {and}
  \bibinfo{person}{Jennifer~Wortman Vaughan}.} \bibinfo{year}{2015}\natexlab{}.
\newblock \showarticletitle{{Incentivizing High Quality Crowdwork}}. In
  \bibinfo{booktitle}{\emph{Proceedings of the 24th International Conference on
  World Wide Web}} \emph{(\bibinfo{series}{WWW '15})}.
  \bibinfo{publisher}{International World Wide Web Conferences Steering
  Committee}, \bibinfo{address}{Republic and Canton of Geneva, Switzerland},
  \bibinfo{pages}{419--429}.
\newblock
\showISBNx{978-1-4503-3469-3}
\urldef\tempurl%
\url{https://doi.org/10.1145/2736277.2741102}
\showDOI{\tempurl}


\bibitem[\protect\citeauthoryear{Horton and Chilton}{Horton and
  Chilton}{2010}]%
        {horton2010labor}
\bibfield{author}{\bibinfo{person}{John~Joseph Horton} {and}
  \bibinfo{person}{Lydia~B. Chilton}.} \bibinfo{year}{2010}\natexlab{}.
\newblock \showarticletitle{The Labor Economics of Paid Crowdsourcing}. In
  \bibinfo{booktitle}{\emph{Proceedings of the 11th ACM Conference on
  Electronic Commerce}} (Cambridge, Massachusetts, USA)
  \emph{(\bibinfo{series}{EC '10})}. \bibinfo{publisher}{ACM},
  \bibinfo{address}{New York, NY, USA}, \bibinfo{pages}{209–218}.
\newblock
\showISBNx{9781605588223}
\urldef\tempurl%
\url{https://doi.org/10.1145/1807342.1807376}
\showDOI{\tempurl}


\bibitem[\protect\citeauthoryear{Huang and Fu}{Huang and Fu}{2013}]%
        {huang2013enhancing}
\bibfield{author}{\bibinfo{person}{Shih-Wen Huang} {and}
  \bibinfo{person}{Wai-Tat Fu}.} \bibinfo{year}{2013}\natexlab{}.
\newblock \showarticletitle{Enhancing Reliability Using Peer Consistency
  Evaluation in Human Computation}. In \bibinfo{booktitle}{\emph{Proceedings of
  the 2013 Conference on Computer Supported Cooperative Work}} (San Antonio,
  Texas, USA) \emph{(\bibinfo{series}{CSCW '13})}. \bibinfo{publisher}{ACM},
  \bibinfo{address}{New York, NY, USA}, \bibinfo{pages}{639–648}.
\newblock
\showISBNx{9781450313315}
\urldef\tempurl%
\url{https://doi.org/10.1145/2441776.2441847}
\showDOI{\tempurl}


\bibitem[\protect\citeauthoryear{Hung, Thang, Weidlich, and Aberer}{Hung
  et~al\mbox{.}}{2015}]%
        {hung2015minimizing}
\bibfield{author}{\bibinfo{person}{Nguyen Quoc~Viet Hung},
  \bibinfo{person}{Duong~Chi Thang}, \bibinfo{person}{Matthias Weidlich}, {and}
  \bibinfo{person}{Karl Aberer}.} \bibinfo{year}{2015}\natexlab{}.
\newblock \showarticletitle{Minimizing Efforts in Validating Crowd Answers}. In
  \bibinfo{booktitle}{\emph{Proceedings of the 2015 ACM SIGMOD International
  Conference on Management of Data}} (Melbourne, Victoria, Australia)
  \emph{(\bibinfo{series}{SIGMOD '15})}. \bibinfo{publisher}{ACM},
  \bibinfo{address}{New York, NY, USA}, \bibinfo{pages}{999–1014}.
\newblock
\showISBNx{9781450327589}
\urldef\tempurl%
\url{https://doi.org/10.1145/2723372.2723731}
\showDOI{\tempurl}


\bibitem[\protect\citeauthoryear{Ipeirotis}{Ipeirotis}{2011}]%
        {ipeirotis-2011}
\bibfield{author}{\bibinfo{person}{Panos Ipeirotis}.}
  \bibinfo{year}{2011}\natexlab{}.
\newblock \bibinfo{title}{{Pay Enough or Don't Pay at All}}.
\newblock
\newblock
\newblock
\shownote{May 13.
  \url{https://www.behind-the-enemy-lines.com/2011/05/pay-enough-or-dont-pay-at-all.html}.}


\bibitem[\protect\citeauthoryear{K.~Chaithanya~Manam, Jampani, Zaim, Wu, and
  J.~Quinn}{K.~Chaithanya~Manam et~al\mbox{.}}{2019}]%
        {Manam2019TaskMate}
\bibfield{author}{\bibinfo{person}{V. K.~Chaithanya~Manam},
  \bibinfo{person}{Dwarakanath Jampani}, \bibinfo{person}{Mariam Zaim},
  \bibinfo{person}{Meng-Han Wu}, {and} \bibinfo{person}{Alexander J.~Quinn}.}
  \bibinfo{year}{2019}\natexlab{}.
\newblock \showarticletitle{TaskMate: A Mechanism to Improve the Quality of
  Instructions in Crowdsourcing}. In \bibinfo{booktitle}{\emph{Companion
  Proceedings of The 2019 World Wide Web Conference}} (San Francisco, USA)
  \emph{(\bibinfo{series}{WWW '19})}. \bibinfo{publisher}{ACM},
  \bibinfo{address}{New York, NY, USA}, \bibinfo{pages}{1121–1130}.
\newblock
\showISBNx{9781450366755}
\urldef\tempurl%
\url{https://doi.org/10.1145/3308560.3317081}
\showDOI{\tempurl}


\bibitem[\protect\citeauthoryear{Kazai}{Kazai}{2011}]%
        {kazai2011search}
\bibfield{author}{\bibinfo{person}{Gabriella Kazai}.}
  \bibinfo{year}{2011}\natexlab{}.
\newblock \showarticletitle{In Search of Quality in Crowdsourcing for Search
  Engine Evaluation}. In \bibinfo{booktitle}{\emph{Advances in Information
  Retrieval}}. \bibinfo{publisher}{Springer}, \bibinfo{address}{Berlin,
  Heidelberg}, \bibinfo{pages}{165--176}.
\newblock
\urldef\tempurl%
\url{https://doi.org/10.1007/978-3-642-20161-5_17}
\showDOI{\tempurl}


\bibitem[\protect\citeauthoryear{Kelling, Gerbracht, Fink, Lagoze, Wong, Yu,
  Damoulas, and Gomes}{Kelling et~al\mbox{.}}{2012}]%
        {kelling2013human}
\bibfield{author}{\bibinfo{person}{Steve Kelling}, \bibinfo{person}{Jeff
  Gerbracht}, \bibinfo{person}{Daniel Fink}, \bibinfo{person}{Carl Lagoze},
  \bibinfo{person}{Weng-Keen Wong}, \bibinfo{person}{Jun Yu},
  \bibinfo{person}{Theodoros Damoulas}, {and} \bibinfo{person}{Carla Gomes}.}
  \bibinfo{year}{2012}\natexlab{}.
\newblock \showarticletitle{A Human/Computer Learning Network to Improve
  Biodiversity Conservation and Research}.
\newblock \bibinfo{journal}{\emph{AI Magazine}} \bibinfo{volume}{34},
  \bibinfo{number}{1} (\bibinfo{date}{Dec.} \bibinfo{year}{2012}),
  \bibinfo{pages}{10}.
\newblock
\urldef\tempurl%
\url{https://doi.org/10.1609/aimag.v34i1.2431}
\showDOI{\tempurl}


\bibitem[\protect\citeauthoryear{Khan and Garcia-Molina}{Khan and
  Garcia-Molina}{2017}]%
        {Khan2017CrowdDQS}
\bibfield{author}{\bibinfo{person}{Asif~R. Khan} {and} \bibinfo{person}{Hector
  Garcia-Molina}.} \bibinfo{year}{2017}\natexlab{}.
\newblock \showarticletitle{{CrowdDQS: Dynamic Question Selection in
  Crowdsourcing Systems}}. In \bibinfo{booktitle}{\emph{Proceedings of the 2017
  ACM International Conference on Management of Data}}
  \emph{(\bibinfo{series}{SIGMOD '17})}. \bibinfo{publisher}{ACM},
  \bibinfo{address}{New York, NY, USA}, \bibinfo{pages}{1447--1462}.
\newblock
\showISBNx{978-1-4503-4197-4}
\urldef\tempurl%
\url{https://doi.org/10.1145/3035918.3064055}
\showDOI{\tempurl}


\bibitem[\protect\citeauthoryear{Kinney, Huffman, and Zhai}{Kinney
  et~al\mbox{.}}{2008}]%
        {Kinney:2008:EDE:1458082.1458160}
\bibfield{author}{\bibinfo{person}{Kenneth~A. Kinney},
  \bibinfo{person}{Scott~B. Huffman}, {and} \bibinfo{person}{Juting Zhai}.}
  \bibinfo{year}{2008}\natexlab{}.
\newblock \showarticletitle{How Evaluator Domain Expertise Affects Search
  Result Relevance Judgments}. In \bibinfo{booktitle}{\emph{Proceedings of the
  17th ACM Conference on Information and Knowledge Management}} (Napa Valley,
  California, USA) \emph{(\bibinfo{series}{CIKM '08})}.
  \bibinfo{publisher}{ACM}, \bibinfo{address}{New York, NY, USA},
  \bibinfo{pages}{591--598}.
\newblock
\showISBNx{978-1-59593-991-3}
\urldef\tempurl%
\url{https://doi.org/10.1145/1458082.1458160}
\showDOI{\tempurl}


\bibitem[\protect\citeauthoryear{Kittur, Khamkar, Andr\'{e}, and Kraut}{Kittur
  et~al\mbox{.}}{2012}]%
        {kittur2012crowdweaver}
\bibfield{author}{\bibinfo{person}{Aniket Kittur}, \bibinfo{person}{Susheel
  Khamkar}, \bibinfo{person}{Paul Andr\'{e}}, {and} \bibinfo{person}{Robert
  Kraut}.} \bibinfo{year}{2012}\natexlab{}.
\newblock \showarticletitle{CrowdWeaver: Visually Managing Complex Crowd Work}.
  In \bibinfo{booktitle}{\emph{Proceedings of the ACM 2012 Conference on
  Computer Supported Cooperative Work}} (Seattle, Washington, USA)
  \emph{(\bibinfo{series}{CSCW '12})}. \bibinfo{publisher}{ACM},
  \bibinfo{address}{New York, NY, USA}, \bibinfo{pages}{1033–1036}.
\newblock
\showISBNx{9781450310864}
\urldef\tempurl%
\url{https://doi.org/10.1145/2145204.2145357}
\showDOI{\tempurl}


\bibitem[\protect\citeauthoryear{Kittur, Smus, Khamkar, and Kraut}{Kittur
  et~al\mbox{.}}{2011}]%
        {kittur2011crowdforge}
\bibfield{author}{\bibinfo{person}{Aniket Kittur}, \bibinfo{person}{Boris
  Smus}, \bibinfo{person}{Susheel Khamkar}, {and} \bibinfo{person}{Robert~E.
  Kraut}.} \bibinfo{year}{2011}\natexlab{}.
\newblock \showarticletitle{CrowdForge: Crowdsourcing Complex Work}. In
  \bibinfo{booktitle}{\emph{Proceedings of the 24th Annual ACM Symposium on
  User Interface Software and Technology}} (Santa Barbara, California, USA)
  \emph{(\bibinfo{series}{UIST '11})}. \bibinfo{publisher}{ACM},
  \bibinfo{address}{New York, NY, USA}, \bibinfo{pages}{43–52}.
\newblock
\showISBNx{9781450307161}
\urldef\tempurl%
\url{https://doi.org/10.1145/2047196.2047202}
\showDOI{\tempurl}


\bibitem[\protect\citeauthoryear{Kovashka, Russakovsky, and Fei-Fei}{Kovashka
  et~al\mbox{.}}{2016}]%
        {Kovashka2016Crowdsourcing}
\bibfield{author}{\bibinfo{person}{Adriana Kovashka}, \bibinfo{person}{Olga
  Russakovsky}, {and} \bibinfo{person}{Li Fei-Fei}.}
  \bibinfo{year}{2016}\natexlab{}.
\newblock \bibinfo{booktitle}{\emph{Crowdsourcing in Computer Vision}}.
\newblock \bibinfo{publisher}{Now Publishers Inc.}, \bibinfo{address}{Hanover,
  MA, USA}.
\newblock
\showISBNx{1680832123}


\bibitem[\protect\citeauthoryear{Kulkarni, Can, and Hartmann}{Kulkarni
  et~al\mbox{.}}{2012}]%
        {Kulkarni2012Collaboratively}
\bibfield{author}{\bibinfo{person}{Anand Kulkarni}, \bibinfo{person}{Matthew
  Can}, {and} \bibinfo{person}{Bj\"{o}rn Hartmann}.}
  \bibinfo{year}{2012}\natexlab{}.
\newblock \showarticletitle{Collaboratively Crowdsourcing Workflows with
  Turkomatic}. In \bibinfo{booktitle}{\emph{Proceedings of the ACM 2012
  Conference on Computer Supported Cooperative Work}} (Seattle, Washington,
  USA) \emph{(\bibinfo{series}{CSCW '12})}. \bibinfo{publisher}{ACM},
  \bibinfo{address}{New York, NY, USA}, \bibinfo{pages}{1003–1012}.
\newblock
\showISBNx{9781450310864}
\urldef\tempurl%
\url{https://doi.org/10.1145/2145204.2145354}
\showDOI{\tempurl}


\bibitem[\protect\citeauthoryear{Kutlu, McDonnell, Elsayed, and Lease}{Kutlu
  et~al\mbox{.}}{2020}]%
        {Kutlu20-jair}
\bibfield{author}{\bibinfo{person}{Mucahid Kutlu}, \bibinfo{person}{Tyler
  McDonnell}, \bibinfo{person}{Tamer Elsayed}, {and} \bibinfo{person}{Matthew
  Lease}.} \bibinfo{year}{2020}\natexlab{}.
\newblock \showarticletitle{{Annotator Rationales for Labeling Tasks in
  Crowdsourcing}}.
\newblock \bibinfo{journal}{\emph{Journal of Artificial Intelligence Research
  (JAIR)}}  \bibinfo{volume}{69} (\bibinfo{year}{2020}),
  \bibinfo{pages}{143--189}.
\newblock
\urldef\tempurl%
\url{https://doi.org/10.1613/jair.1.12012}
\showDOI{\tempurl}


\bibitem[\protect\citeauthoryear{Kuznetsova, Rom, Alldrin, Uijlings, Krasin,
  Pont-Tuset, Kamali, Popov, Malloci, Kolesnikov, Duerig, and
  Ferrari}{Kuznetsova et~al\mbox{.}}{2020}]%
        {OpenImages}
\bibfield{author}{\bibinfo{person}{Alina Kuznetsova}, \bibinfo{person}{Hassan
  Rom}, \bibinfo{person}{Neil Alldrin}, \bibinfo{person}{Jasper Uijlings},
  \bibinfo{person}{Ivan Krasin}, \bibinfo{person}{Jordi Pont-Tuset},
  \bibinfo{person}{Shahab Kamali}, \bibinfo{person}{Stefan Popov},
  \bibinfo{person}{Matteo Malloci}, \bibinfo{person}{Alexander Kolesnikov},
  \bibinfo{person}{Tom Duerig}, {and} \bibinfo{person}{Vittorio Ferrari}.}
  \bibinfo{year}{2020}\natexlab{}.
\newblock \showarticletitle{The Open Images Dataset V4: Unified image
  classification, object detection, and visual relationship detection at
  scale}.
\newblock \bibinfo{journal}{\emph{International Journal of Computer Vision}}
  \bibinfo{volume}{128}, \bibinfo{number}{7} (\bibinfo{year}{2020}),
  \bibinfo{pages}{1956--1981}.
\newblock
\urldef\tempurl%
\url{https://doi.org/10.1007/s11263-020-01316-z}
\showDOI{\tempurl}


\bibitem[\protect\citeauthoryear{Le, Edmonds, Hester, and Biewald}{Le
  et~al\mbox{.}}{2010}]%
        {le2010ensuring}
\bibfield{author}{\bibinfo{person}{John Le}, \bibinfo{person}{Andy Edmonds},
  \bibinfo{person}{Vaughn Hester}, {and} \bibinfo{person}{Lukas Biewald}.}
  \bibinfo{year}{2010}\natexlab{}.
\newblock \showarticletitle{Ensuring quality in crowdsourced search relevance
  evaluation: The effects of training question distribution}. In
  \bibinfo{booktitle}{\emph{SIGIR 2010 Workshop on Crowdsourcing for Search
  Evaluation}}. \bibinfo{pages}{21--26}.
\newblock


\bibitem[\protect\citeauthoryear{Lintott, Schawinski, Slosar, Land, Bamford,
  Thomas, Raddick, Nichol, Szalay, Andreescu, et~al\mbox{.}}{Lintott
  et~al\mbox{.}}{2008}]%
        {lintott2008galaxy}
\bibfield{author}{\bibinfo{person}{Chris~J. Lintott}, \bibinfo{person}{Kevin
  Schawinski}, \bibinfo{person}{An{\v{z}}e Slosar}, \bibinfo{person}{Kate
  Land}, \bibinfo{person}{Steven Bamford}, \bibinfo{person}{Daniel Thomas},
  \bibinfo{person}{M.~Jordan Raddick}, \bibinfo{person}{Robert~C. Nichol},
  \bibinfo{person}{Alex Szalay}, \bibinfo{person}{Dan Andreescu},
  {et~al\mbox{.}}} \bibinfo{year}{2008}\natexlab{}.
\newblock \showarticletitle{Galaxy Zoo: morphologies derived from visual
  inspection of galaxies from the Sloan Digital Sky Survey}.
\newblock \bibinfo{journal}{\emph{Monthly Notices of the Royal Astronomical
  Society}} \bibinfo{volume}{389}, \bibinfo{number}{3} (\bibinfo{year}{2008}),
  \bibinfo{pages}{1179--1189}.
\newblock


\bibitem[\protect\citeauthoryear{Little, Chilton, Goldman, and Miller}{Little
  et~al\mbox{.}}{2010}]%
        {little2010exploring}
\bibfield{author}{\bibinfo{person}{Greg Little}, \bibinfo{person}{Lydia~B.
  Chilton}, \bibinfo{person}{Max Goldman}, {and} \bibinfo{person}{Robert~C.
  Miller}.} \bibinfo{year}{2010}\natexlab{}.
\newblock \showarticletitle{Exploring iterative and parallel human computation
  processes}. In \bibinfo{booktitle}{\emph{Proceedings of the ACM SIGKDD
  workshop on human computation}}. \bibinfo{pages}{68--76}.
\newblock


\bibitem[\protect\citeauthoryear{Liu, Ouyang, Wang, Fieguth, Chen, Liu, and
  Pietik{\"a}inen}{Liu et~al\mbox{.}}{2020}]%
        {liu2020deep}
\bibfield{author}{\bibinfo{person}{Li Liu}, \bibinfo{person}{Wanli Ouyang},
  \bibinfo{person}{Xiaogang Wang}, \bibinfo{person}{Paul Fieguth},
  \bibinfo{person}{Jie Chen}, \bibinfo{person}{Xinwang Liu}, {and}
  \bibinfo{person}{Matti Pietik{\"a}inen}.} \bibinfo{year}{2020}\natexlab{}.
\newblock \showarticletitle{Deep learning for generic object detection: A
  survey}.
\newblock \bibinfo{journal}{\emph{International journal of computer vision}}
  \bibinfo{volume}{128}, \bibinfo{number}{2} (\bibinfo{year}{2020}),
  \bibinfo{pages}{261--318}.
\newblock
\urldef\tempurl%
\url{https://doi.org/10.1007/s11263-019-01247-4}
\showDOI{\tempurl}


\bibitem[\protect\citeauthoryear{Liu, Ihler, and Steyvers}{Liu
  et~al\mbox{.}}{2013}]%
        {liu2013scoring}
\bibfield{author}{\bibinfo{person}{Qiang Liu}, \bibinfo{person}{Alexander~T
  Ihler}, {and} \bibinfo{person}{Mark Steyvers}.}
  \bibinfo{year}{2013}\natexlab{}.
\newblock \showarticletitle{Scoring Workers in Crowdsourcing: How Many Control
  Questions are Enough?}. In \bibinfo{booktitle}{\emph{Advances in Neural
  Information Processing Systems}}, Vol.~\bibinfo{volume}{26}.
  \bibinfo{publisher}{Curran Associates, Inc.}, \bibinfo{pages}{1914--1922}.
\newblock


\bibitem[\protect\citeauthoryear{Manam and Quinn}{Manam and Quinn}{2018}]%
        {manam2018wingit}
\bibfield{author}{\bibinfo{person}{V.K.~Chaithanya Manam} {and}
  \bibinfo{person}{Alexander~J. Quinn}.} \bibinfo{year}{2018}\natexlab{}.
\newblock \showarticletitle{WingIt: Efficient refinement of unclear task
  instructions}. In \bibinfo{booktitle}{\emph{Sixth AAAI Conference on Human
  Computation and Crowdsourcing}} \emph{(\bibinfo{series}{HCOMP},
  Vol.~\bibinfo{volume}{6})}. \bibinfo{publisher}{AAAI Press}.
\newblock


\bibitem[\protect\citeauthoryear{Mankar, Shah, and Lease}{Mankar
  et~al\mbox{.}}{2017}]%
        {Mankar17-hcomp}
\bibfield{author}{\bibinfo{person}{Akash Mankar}, \bibinfo{person}{Riddhi~J.\
  Shah}, {and} \bibinfo{person}{Matthew Lease}.}
  \bibinfo{year}{2017}\natexlab{}.
\newblock \showarticletitle{{Design Activism for Minimum Wage Crowd Work}}. In
  \bibinfo{booktitle}{\emph{Fifth AAAI Conference on Human Computation and
  Crowdsourcing (HCOMP): Works-in-Progress Track}}.
\newblock


\bibitem[\protect\citeauthoryear{Mason and Watts}{Mason and Watts}{2009}]%
        {mason2009financial}
\bibfield{author}{\bibinfo{person}{Winter Mason} {and}
  \bibinfo{person}{Duncan~J. Watts}.} \bibinfo{year}{2009}\natexlab{}.
\newblock \showarticletitle{Financial incentives and the performance of
  crowds}. In \bibinfo{booktitle}{\emph{Proceedings of the ACM SIGKDD Workshop
  on Human Computation}}. ACM, \bibinfo{pages}{77--85}.
\newblock


\bibitem[\protect\citeauthoryear{Newell and Ruths}{Newell and Ruths}{2016}]%
        {Newell2016}
\bibfield{author}{\bibinfo{person}{Edward Newell} {and} \bibinfo{person}{Derek
  Ruths}.} \bibinfo{year}{2016}\natexlab{}.
\newblock \showarticletitle{{How One Microtask Affects Another}}. In
  \bibinfo{booktitle}{\emph{Proceedings of the 2016 CHI Conference on Human
  Factors in Computing Systems}} \emph{(\bibinfo{series}{CHI '16})}.
  \bibinfo{publisher}{ACM}, \bibinfo{address}{New York, NY, USA},
  \bibinfo{pages}{3155–3166}.
\newblock
\showISBNx{9781450333627}
\urldef\tempurl%
\url{https://doi.org/10.1145/2858036.2858490}
\showDOI{\tempurl}


\bibitem[\protect\citeauthoryear{Oleson, Sorokin, Laughlin, Hester, Le, and
  Biewald}{Oleson et~al\mbox{.}}{2011}]%
        {oleson2011programmatic}
\bibfield{author}{\bibinfo{person}{David Oleson}, \bibinfo{person}{Alexander
  Sorokin}, \bibinfo{person}{Greg Laughlin}, \bibinfo{person}{Vaughn Hester},
  \bibinfo{person}{John Le}, {and} \bibinfo{person}{Lukas Biewald}.}
  \bibinfo{year}{2011}\natexlab{}.
\newblock \showarticletitle{Programmatic gold: Targeted and scalable quality
  assurance in crowdsourcing}. In \bibinfo{booktitle}{\emph{Proceedings of the
  AAAI Conference on Human Computation}} \emph{(\bibinfo{series}{HCOMP})}. AAAI
  Press.
\newblock


\bibitem[\protect\citeauthoryear{Papadopoulos, Uijlings, Keller, and
  Ferrari}{Papadopoulos et~al\mbox{.}}{2016}]%
        {papadopoulos2016we}
\bibfield{author}{\bibinfo{person}{Dim~P. Papadopoulos},
  \bibinfo{person}{Jasper~R.R. Uijlings}, \bibinfo{person}{Frank Keller}, {and}
  \bibinfo{person}{Vittorio Ferrari}.} \bibinfo{year}{2016}\natexlab{}.
\newblock \showarticletitle{We Don’t Need No Bounding-Boxes: Training Object
  Class Detectors Using Only Human Verification}. In
  \bibinfo{booktitle}{\emph{2016 IEEE Conference on Computer Vision and Pattern
  Recognition (CVPR)}}. \bibinfo{pages}{854--863}.
\newblock
\urldef\tempurl%
\url{https://doi.org/10.1109/CVPR.2016.99}
\showDOI{\tempurl}


\bibitem[\protect\citeauthoryear{Papadopoulos, Uijlings, Keller, and
  Ferrari}{Papadopoulos et~al\mbox{.}}{2017a}]%
        {papadopoulos2017extreme}
\bibfield{author}{\bibinfo{person}{Dim~P. Papadopoulos},
  \bibinfo{person}{Jasper~R.R. Uijlings}, \bibinfo{person}{Frank Keller}, {and}
  \bibinfo{person}{Vittorio Ferrari}.} \bibinfo{year}{2017}\natexlab{a}.
\newblock \showarticletitle{Extreme Clicking for Efficient Object Annotation}.
  In \bibinfo{booktitle}{\emph{2017 IEEE International Conference on Computer
  Vision (ICCV)}}. \bibinfo{pages}{4940--4949}.
\newblock
\urldef\tempurl%
\url{https://doi.org/10.1109/ICCV.2017.528}
\showDOI{\tempurl}


\bibitem[\protect\citeauthoryear{Papadopoulos, Uijlings, Keller, and
  Ferrari}{Papadopoulos et~al\mbox{.}}{2017b}]%
        {papadopoulos2017training}
\bibfield{author}{\bibinfo{person}{Dim~P. Papadopoulos},
  \bibinfo{person}{Jasper~R.R. Uijlings}, \bibinfo{person}{Frank Keller}, {and}
  \bibinfo{person}{Vittorio Ferrari}.} \bibinfo{year}{2017}\natexlab{b}.
\newblock \showarticletitle{Training Object Class Detectors with Click
  Supervision}. In \bibinfo{booktitle}{\emph{2017 IEEE Conference on Computer
  Vision and Pattern Recognition (CVPR)}}. \bibinfo{pages}{180--189}.
\newblock
\urldef\tempurl%
\url{https://doi.org/10.1109/CVPR.2017.27}
\showDOI{\tempurl}


\bibitem[\protect\citeauthoryear{Park, Shoemark, and Morency}{Park
  et~al\mbox{.}}{2014}]%
        {park2014toward}
\bibfield{author}{\bibinfo{person}{Sunghyun Park}, \bibinfo{person}{Philippa
  Shoemark}, {and} \bibinfo{person}{Louis-Philippe Morency}.}
  \bibinfo{year}{2014}\natexlab{}.
\newblock \showarticletitle{Toward Crowdsourcing Micro-Level Behavior
  Annotations: The Challenges of Interface, Training, and Generalization}. In
  \bibinfo{booktitle}{\emph{Proceedings of the 19th International Conference on
  Intelligent User Interfaces}} (Haifa, Israel) \emph{(\bibinfo{series}{IUI
  '14})}. \bibinfo{publisher}{ACM}, \bibinfo{address}{New York, NY, USA},
  \bibinfo{pages}{37–46}.
\newblock
\showISBNx{9781450321846}
\urldef\tempurl%
\url{https://doi.org/10.1145/2557500.2557512}
\showDOI{\tempurl}


\bibitem[\protect\citeauthoryear{Retelny, Bernstein, and Valentine}{Retelny
  et~al\mbox{.}}{2017}]%
        {retelny2017no}
\bibfield{author}{\bibinfo{person}{Daniela Retelny},
  \bibinfo{person}{Michael~S. Bernstein}, {and} \bibinfo{person}{Melissa~A.
  Valentine}.} \bibinfo{year}{2017}\natexlab{}.
\newblock \showarticletitle{No Workflow Can Ever Be Enough: How Crowdsourcing
  Workflows Constrain Complex Work}.
\newblock \bibinfo{journal}{\emph{Proc. ACM Hum.-Comput. Interact.}}
  \bibinfo{volume}{1}, \bibinfo{number}{CSCW}, Article \bibinfo{articleno}{89}
  (\bibinfo{date}{Dec.} \bibinfo{year}{2017}), \bibinfo{numpages}{23}~pages.
\newblock
\urldef\tempurl%
\url{https://doi.org/10.1145/3134724}
\showDOI{\tempurl}


\bibitem[\protect\citeauthoryear{Rogstadius, Kostakos, Kittur, Smus, Laredo,
  and Vukovic}{Rogstadius et~al\mbox{.}}{2011}]%
        {rogstadius2011assessment}
\bibfield{author}{\bibinfo{person}{Jakob Rogstadius}, \bibinfo{person}{Vassilis
  Kostakos}, \bibinfo{person}{Aniket Kittur}, \bibinfo{person}{Boris Smus},
  \bibinfo{person}{Jim Laredo}, {and} \bibinfo{person}{Maja Vukovic}.}
  \bibinfo{year}{2011}\natexlab{}.
\newblock \showarticletitle{An assessment of intrinsic and extrinsic motivation
  on task performance in crowdsourcing markets}. In
  \bibinfo{booktitle}{\emph{Proceedings of the International AAAI Conference on
  Web and Social Media}}, Vol.~\bibinfo{volume}{5}. \bibinfo{publisher}{AAAI
  Press}.
\newblock


\bibitem[\protect\citeauthoryear{Rosser and Wiggins}{Rosser and
  Wiggins}{2019}]%
        {rosser2019crowds}
\bibfield{author}{\bibinfo{person}{Holly Rosser} {and} \bibinfo{person}{Andrea
  Wiggins}.} \bibinfo{year}{2019}\natexlab{}.
\newblock \showarticletitle{Crowds and Camera Traps: Genres in Online Citizen
  Science Projects}. In \bibinfo{booktitle}{\emph{Proceedings of the 52nd
  Hawaii International Conference on System Sciences}}.
\newblock


\bibitem[\protect\citeauthoryear{Russakovsky, Li, and Fei-Fei}{Russakovsky
  et~al\mbox{.}}{2015}]%
        {russakovsky2015best}
\bibfield{author}{\bibinfo{person}{Olga Russakovsky}, \bibinfo{person}{Li-Jia
  Li}, {and} \bibinfo{person}{Li Fei-Fei}.} \bibinfo{year}{2015}\natexlab{}.
\newblock \showarticletitle{Best of both worlds: human-machine collaboration
  for object annotation}. In \bibinfo{booktitle}{\emph{2015 IEEE Conference on
  Computer Vision and Pattern Recognition (CVPR)}}.
  \bibinfo{pages}{2121--2131}.
\newblock
\urldef\tempurl%
\url{https://doi.org/10.1109/CVPR.2015.7298824}
\showDOI{\tempurl}


\bibitem[\protect\citeauthoryear{Sarma, Jain, Nandi, Parameswaran, and
  Widom}{Sarma et~al\mbox{.}}{2015}]%
        {sarma2015surpassing}
\bibfield{author}{\bibinfo{person}{Akash Sarma}, \bibinfo{person}{Ayush Jain},
  \bibinfo{person}{Arnab Nandi}, \bibinfo{person}{Aditya Parameswaran}, {and}
  \bibinfo{person}{Jennifer Widom}.} \bibinfo{year}{2015}\natexlab{}.
\newblock \showarticletitle{Surpassing humans and computers with jellybean:
  Crowd-vision-hybrid counting algorithms}. In
  \bibinfo{booktitle}{\emph{Proceedings of the Third AAAI Conference on Human
  Computation and Crowdsourcing}} \emph{(\bibinfo{series}{HCOMP},
  Vol.~\bibinfo{volume}{3})}. \bibinfo{publisher}{AAAI Press}.
\newblock


\bibitem[\protect\citeauthoryear{Savage, Chiang, Saito, Toxtli, and
  Bigham}{Savage et~al\mbox{.}}{2020}]%
        {Savage2020BecomingWorkers}
\bibfield{author}{\bibinfo{person}{Saiph Savage}, \bibinfo{person}{Chun~Wei
  Chiang}, \bibinfo{person}{Susumu Saito}, \bibinfo{person}{Carlos Toxtli},
  {and} \bibinfo{person}{Jeffrey Bigham}.} \bibinfo{year}{2020}\natexlab{}.
\newblock \showarticletitle{{Becoming the Super Turker:Increasing Wages via a
  Strategy from High Earning Workers}}. In
  \bibinfo{booktitle}{\emph{Proceedings of The Web Conference 2020}}
  \emph{(\bibinfo{series}{WWW '20})}. \bibinfo{publisher}{ACM},
  \bibinfo{address}{New York, NY, USA}, \bibinfo{pages}{1241–1252}.
\newblock
\showISBNx{9781450370233}
\urldef\tempurl%
\url{https://doi.org/10.1145/3366423.3380200}
\showDOI{\tempurl}


\bibitem[\protect\citeauthoryear{Schaekermann, Beaton, Sanoubari, Lim, Larson,
  and Law}{Schaekermann et~al\mbox{.}}{2020}]%
        {Schaekermann2020}
\bibfield{author}{\bibinfo{person}{Mike Schaekermann}, \bibinfo{person}{Graeme
  Beaton}, \bibinfo{person}{Elaheh Sanoubari}, \bibinfo{person}{Andrew Lim},
  \bibinfo{person}{Kate Larson}, {and} \bibinfo{person}{Edith Law}.}
  \bibinfo{year}{2020}\natexlab{}.
\newblock \showarticletitle{Ambiguity-Aware AI Assistants for Medical Data
  Analysis}. In \bibinfo{booktitle}{\emph{Proceedings of the 2020 CHI
  Conference on Human Factors in Computing Systems}} (Honolulu, HI, USA)
  \emph{(\bibinfo{series}{CHI '20})}. \bibinfo{publisher}{ACM},
  \bibinfo{address}{New York, NY, USA}, \bibinfo{pages}{1–14}.
\newblock
\showISBNx{9781450367080}
\urldef\tempurl%
\url{https://doi.org/10.1145/3313831.3376506}
\showDOI{\tempurl}


\bibitem[\protect\citeauthoryear{Schaekermann, Goh, Larson, and
  Law}{Schaekermann et~al\mbox{.}}{2018}]%
        {Schaekermann2018}
\bibfield{author}{\bibinfo{person}{Mike Schaekermann}, \bibinfo{person}{Joslin
  Goh}, \bibinfo{person}{Kate Larson}, {and} \bibinfo{person}{Edith Law}.}
  \bibinfo{year}{2018}\natexlab{}.
\newblock \showarticletitle{Resolvable vs. Irresolvable Disagreement: A Study
  on Worker Deliberation in Crowd Work}.
\newblock \bibinfo{journal}{\emph{Proc. ACM Hum.-Comput. Interact.}}
  \bibinfo{volume}{2}, \bibinfo{number}{CSCW}, Article \bibinfo{articleno}{154}
  (\bibinfo{date}{Nov.} \bibinfo{year}{2018}), \bibinfo{numpages}{19}~pages.
\newblock
\urldef\tempurl%
\url{https://doi.org/10.1145/3274423}
\showDOI{\tempurl}


\bibitem[\protect\citeauthoryear{SetiHome}{SetiHome}{2021}]%
        {setiathome}
\bibfield{author}{\bibinfo{person}{SetiHome}.} \bibinfo{year}{2021}\natexlab{}.
\newblock \bibinfo{title}{SetiHome}.
\newblock
\newblock
\urldef\tempurl%
\url{https://setiathome.berkeley.edu}
\showURL{%
\tempurl}


\bibitem[\protect\citeauthoryear{Shao, Li, Zhang, Peng, Yu, Zhang, Li, and
  Sun}{Shao et~al\mbox{.}}{2019}]%
        {shao2019objects365}
\bibfield{author}{\bibinfo{person}{Shuai Shao}, \bibinfo{person}{Zeming Li},
  \bibinfo{person}{Tianyuan Zhang}, \bibinfo{person}{Chao Peng},
  \bibinfo{person}{Gang Yu}, \bibinfo{person}{Xiangyu Zhang},
  \bibinfo{person}{Jing Li}, {and} \bibinfo{person}{Jian Sun}.}
  \bibinfo{year}{2019}\natexlab{}.
\newblock \showarticletitle{Objects365: A large-scale, high-quality dataset for
  object detection}. In \bibinfo{booktitle}{\emph{Proceedings of the IEEE
  international conference on computer vision}}. \bibinfo{pages}{8430--8439}.
\newblock


\bibitem[\protect\citeauthoryear{Su, Deng, and Fei-Fei}{Su
  et~al\mbox{.}}{2012}]%
        {su2012crowdsourcing}
\bibfield{author}{\bibinfo{person}{Hao Su}, \bibinfo{person}{Jia Deng}, {and}
  \bibinfo{person}{Li Fei-Fei}.} \bibinfo{year}{2012}\natexlab{}.
\newblock \showarticletitle{Crowdsourcing Annotations for Visual Object
  Detection}. In \bibinfo{booktitle}{\emph{Workshops at the Twenty-Sixth AAAI
  Conference on Artificial Intelligence}}. \bibinfo{publisher}{AAAI Press}.
\newblock


\bibitem[\protect\citeauthoryear{Tang, Cebrian, Giacobe, Kim, Kim, and
  Wickert}{Tang et~al\mbox{.}}{2011}]%
        {tang2011reflecting}
\bibfield{author}{\bibinfo{person}{John~C Tang}, \bibinfo{person}{Manuel
  Cebrian}, \bibinfo{person}{Nicklaus~A Giacobe}, \bibinfo{person}{Hyun-Woo
  Kim}, \bibinfo{person}{Taemie Kim}, {and} \bibinfo{person}{Douglas~Beaker
  Wickert}.} \bibinfo{year}{2011}\natexlab{}.
\newblock \showarticletitle{Reflecting on the DARPA red balloon challenge}.
\newblock \bibinfo{journal}{\emph{Commun. ACM}} \bibinfo{volume}{54},
  \bibinfo{number}{4} (\bibinfo{year}{2011}), \bibinfo{pages}{78--85}.
\newblock


\bibitem[\protect\citeauthoryear{Teevan, Iqbal, and von Veh}{Teevan
  et~al\mbox{.}}{2016}]%
        {teevan2016supporting}
\bibfield{author}{\bibinfo{person}{Jaime Teevan}, \bibinfo{person}{Shamsi~T.
  Iqbal}, {and} \bibinfo{person}{Curtis von Veh}.}
  \bibinfo{year}{2016}\natexlab{}.
\newblock \showarticletitle{Supporting Collaborative Writing with Microtasks}.
  In \bibinfo{booktitle}{\emph{Proceedings of the 2016 CHI Conference on Human
  Factors in Computing Systems}} (San Jose, California, USA)
  \emph{(\bibinfo{series}{CHI '16})}. \bibinfo{publisher}{ACM},
  \bibinfo{address}{New York, NY, USA}, \bibinfo{pages}{2657–2668}.
\newblock
\showISBNx{9781450333627}
\urldef\tempurl%
\url{https://doi.org/10.1145/2858036.2858108}
\showDOI{\tempurl}


\bibitem[\protect\citeauthoryear{Tong, Chen, Zhou, Jagadish, Shou, and Lv}{Tong
  et~al\mbox{.}}{2018}]%
        {Tong2018SLADE:Crowdsourcing}
\bibfield{author}{\bibinfo{person}{Yongxin Tong}, \bibinfo{person}{Lei Chen},
  \bibinfo{person}{Zimu Zhou}, \bibinfo{person}{H.~V. Jagadish},
  \bibinfo{person}{Lidan Shou}, {and} \bibinfo{person}{Weifeng Lv}.}
  \bibinfo{year}{2018}\natexlab{}.
\newblock \showarticletitle{{SLADE: A Smart Large-Scale Task Decomposer in
  Crowdsourcing}}.
\newblock \bibinfo{journal}{\emph{IEEE Transactions on Knowledge and Data
  Engineering}} \bibinfo{volume}{30}, \bibinfo{number}{8} (\bibinfo{date}{8}
  \bibinfo{year}{2018}), \bibinfo{pages}{1588--1601}.
\newblock
\showISSN{10414347}
\urldef\tempurl%
\url{https://doi.org/10.1109/TKDE.2018.2797962}
\showDOI{\tempurl}


\bibitem[\protect\citeauthoryear{Vaughan}{Vaughan}{2017}]%
        {vaughan2017making}
\bibfield{author}{\bibinfo{person}{Jennifer~Wortman Vaughan}.}
  \bibinfo{year}{2017}\natexlab{}.
\newblock \showarticletitle{Making better use of the crowd: How crowdsourcing
  can advance machine learning research}.
\newblock \bibinfo{journal}{\emph{The Journal of Machine Learning Research}}
  \bibinfo{volume}{18}, \bibinfo{number}{1} (\bibinfo{year}{2017}),
  \bibinfo{pages}{7026--7071}.
\newblock


\bibitem[\protect\citeauthoryear{Vogels}{Vogels}{2007}]%
        {jimgray}
\bibfield{author}{\bibinfo{person}{Werner Vogels}.}
  \bibinfo{year}{2007}\natexlab{}.
\newblock \bibinfo{booktitle}{\emph{Help Find Jim Gray}}.
\newblock
\newblock
\shownote{\url{https://www.allthingsdistributed.com/2007/02/help_find_jim_gray.html}.}


\bibitem[\protect\citeauthoryear{Wang, Hou, Yao, and Yan}{Wang
  et~al\mbox{.}}{2009}]%
        {wang2009human}
\bibfield{author}{\bibinfo{person}{Bing Wang}, \bibinfo{person}{Bonan Hou},
  \bibinfo{person}{Yiping Yao}, {and} \bibinfo{person}{Laibin Yan}.}
  \bibinfo{year}{2009}\natexlab{}.
\newblock \showarticletitle{Human flesh search model incorporating network
  expansion and gossip with feedback}. In \bibinfo{booktitle}{\emph{2009 13th
  IEEE/ACM International Symposium on Distributed Simulation and Real Time
  Applications}}. IEEE, \bibinfo{pages}{82--88}.
\newblock
\urldef\tempurl%
\url{https://doi.org/10.1109/DS-RT.2009.36}
\showDOI{\tempurl}


\bibitem[\protect\citeauthoryear{Whiting, Gamage, Gaikwad, Gilbee, Goyal,
  Ballav, Majeti, Chhibber, Richmond-Fuller, Vargus, Sarma, Chandrakanthan,
  Moura, Salih, Bayomi Tinoco~Kalejaiye, Ginzberg, Mullings, Dayan, Milland,
  Orefice, Regino, Parsi, Mainali, Sehgal, Matin, Sinha, Vaish, and
  Bernstein}{Whiting et~al\mbox{.}}{2017}]%
        {whiting2017crowd}
\bibfield{author}{\bibinfo{person}{Mark~E. Whiting}, \bibinfo{person}{Dilrukshi
  Gamage}, \bibinfo{person}{Snehalkumar (Neil)~S. Gaikwad},
  \bibinfo{person}{Aaron Gilbee}, \bibinfo{person}{Shirish Goyal},
  \bibinfo{person}{Alipta Ballav}, \bibinfo{person}{Dinesh Majeti},
  \bibinfo{person}{Nalin Chhibber}, \bibinfo{person}{Angela Richmond-Fuller},
  \bibinfo{person}{Freddie Vargus}, \bibinfo{person}{Tejas~Seshadri Sarma},
  \bibinfo{person}{Varshine Chandrakanthan}, \bibinfo{person}{Teogenes Moura},
  \bibinfo{person}{Mohamed~Hashim Salih}, \bibinfo{person}{Gabriel Bayomi
  Tinoco~Kalejaiye}, \bibinfo{person}{Adam Ginzberg},
  \bibinfo{person}{Catherine~A. Mullings}, \bibinfo{person}{Yoni Dayan},
  \bibinfo{person}{Kristy Milland}, \bibinfo{person}{Henrique Orefice},
  \bibinfo{person}{Jeff Regino}, \bibinfo{person}{Sayna Parsi},
  \bibinfo{person}{Kunz Mainali}, \bibinfo{person}{Vibhor Sehgal},
  \bibinfo{person}{Sekandar Matin}, \bibinfo{person}{Akshansh Sinha},
  \bibinfo{person}{Rajan Vaish}, {and} \bibinfo{person}{Michael~S. Bernstein}.}
  \bibinfo{year}{2017}\natexlab{}.
\newblock \showarticletitle{Crowd Guilds: Worker-Led Reputation and Feedback on
  Crowdsourcing Platforms}. In \bibinfo{booktitle}{\emph{Proceedings of the
  2017 ACM Conference on Computer Supported Cooperative Work and Social
  Computing}} (Portland, Oregon, USA) \emph{(\bibinfo{series}{CSCW '17})}.
  \bibinfo{publisher}{ACM}, \bibinfo{address}{New York, NY, USA},
  \bibinfo{pages}{1902–1913}.
\newblock
\showISBNx{9781450343350}
\urldef\tempurl%
\url{https://doi.org/10.1145/2998181.2998234}
\showDOI{\tempurl}


\bibitem[\protect\citeauthoryear{Whiting, Hugh, and Bernstein}{Whiting
  et~al\mbox{.}}{2019}]%
        {whiting2019fair}
\bibfield{author}{\bibinfo{person}{Mark~E. Whiting}, \bibinfo{person}{Grant
  Hugh}, {and} \bibinfo{person}{Michael~S. Bernstein}.}
  \bibinfo{year}{2019}\natexlab{}.
\newblock \showarticletitle{Fair Work: Crowd Work Minimum Wage with One Line of
  Code}. In \bibinfo{booktitle}{\emph{Proceedings of the Seventh AAAI
  Conference on Human Computation and Crowdsourcing}}
  \emph{(\bibinfo{series}{HCOMP}, Vol.~\bibinfo{volume}{7})}.
  \bibinfo{pages}{197--206}.
\newblock


\bibitem[\protect\citeauthoryear{Wikipedia}{Wikipedia}{2020}]%
        {waldo}
\bibfield{author}{\bibinfo{person}{Wikipedia}.}
  \bibinfo{year}{2020}\natexlab{}.
\newblock \bibinfo{title}{{Where's Wally?}}
\newblock
\newblock
\newblock
\shownote{\waldourl.}


\bibitem[\protect\citeauthoryear{Yang, Redi, Demartini, and Bozzon}{Yang
  et~al\mbox{.}}{2016}]%
        {Yang2016ModelingCrowdsourcing}
\bibfield{author}{\bibinfo{person}{Jie Yang}, \bibinfo{person}{Judith Redi},
  \bibinfo{person}{Gianluca Demartini}, {and} \bibinfo{person}{Alessandro
  Bozzon}.} \bibinfo{year}{2016}\natexlab{}.
\newblock \showarticletitle{{Modeling Task Complexity in Crowdsourcing}}.
\newblock \bibinfo{journal}{\emph{The Fourth AAAI Conference on Human
  Computation and Crowdsourcing}} \bibinfo{number}{October},
  \bibinfo{pages}{249--258}.
\newblock
\urldef\tempurl%
\url{https://aaai.org/ocs/index.php/HCOMP/HCOMP16/paper/viewFile/14039/13653}
\showURL{%
\tempurl}


\bibitem[\protect\citeauthoryear{Ye, You, and Robert~Jr.}{Ye
  et~al\mbox{.}}{2017}]%
        {Ye2017}
\bibfield{author}{\bibinfo{person}{Teng Ye}, \bibinfo{person}{Sangseok You},
  {and} \bibinfo{person}{Lionel Robert~Jr.}} \bibinfo{year}{2017}\natexlab{}.
\newblock \showarticletitle{When Does More Money Work? Examining the Role of
  Perceived Fairness in Pay on the Performance Quality of Crowdworkers}.
\newblock \bibinfo{journal}{\emph{Proceedings of the International AAAI
  Conference on Web and Social Media}} \bibinfo{volume}{11},
  \bibinfo{number}{1}.
\newblock


\bibitem[\protect\citeauthoryear{Yin and Chen}{Yin and Chen}{2015}]%
        {yin2015bonus}
\bibfield{author}{\bibinfo{person}{Ming Yin} {and} \bibinfo{person}{Yiling
  Chen}.} \bibinfo{year}{2015}\natexlab{}.
\newblock \showarticletitle{Bonus or Not? Learn to Reward in Crowdsourcing}. In
  \bibinfo{booktitle}{\emph{Proceedings of the 24th International Conference on
  Artificial Intelligence}} \emph{(\bibinfo{series}{IJCAI'15})}.
  \bibinfo{publisher}{AAAI Press}, \bibinfo{pages}{201–207}.
\newblock
\showISBNx{9781577357384}


\bibitem[\protect\citeauthoryear{Yin and Chen}{Yin and Chen}{2016}]%
        {yin2016predicting}
\bibfield{author}{\bibinfo{person}{Ming Yin} {and} \bibinfo{person}{Yiling
  Chen}.} \bibinfo{year}{2016}\natexlab{}.
\newblock \showarticletitle{Predicting crowd work quality under monetary
  interventions}. In \bibinfo{booktitle}{\emph{Proceedings of the AAAI
  Conference on Human Computation and Crowdsourcing}}
  \emph{(\bibinfo{series}{HCOMP}, Vol.~\bibinfo{volume}{4})}.
  \bibinfo{publisher}{AAAI Press}.
\newblock


\bibitem[\protect\citeauthoryear{Zhu, Dow, Kraut, and Kittur}{Zhu
  et~al\mbox{.}}{2014}]%
        {Zhu2014}
\bibfield{author}{\bibinfo{person}{Haiyi Zhu}, \bibinfo{person}{Steven~P. Dow},
  \bibinfo{person}{Robert~E. Kraut}, {and} \bibinfo{person}{Aniket Kittur}.}
  \bibinfo{year}{2014}\natexlab{}.
\newblock \showarticletitle{{Reviewing Versus Doing: Learning and Performance
  in Crowd Assessment}}. In \bibinfo{booktitle}{\emph{Proceedings of the 17th
  ACM Conference on Computer Supported Cooperative Work {\&} Social Computing}}
  \emph{(\bibinfo{series}{CSCW '14})}. \bibinfo{publisher}{ACM},
  \bibinfo{address}{New York, NY, USA}, \bibinfo{pages}{1445--1455}.
\newblock
\showISBNx{978-1-4503-2540-0}
\urldef\tempurl%
\url{https://doi.org/10.1145/2531602.2531718}
\showDOI{\tempurl}


\end{thebibliography}



\end{document}